\documentclass[12pt]{article}
\usepackage{amsmath}
\usepackage{commath}
\usepackage{amsthm}
\usepackage{amssymb}
\usepackage{setspace,mathrsfs}
\usepackage{algorithm,natbib}
\usepackage{algorithmic}
\usepackage{url,amsmath,amsfonts,amssymb}
\usepackage{caption,subcaption,algorithm,graphicx}
\usepackage{listings,enumerate,color,relsize,multirow,rotating,soul}

\def\nk{\scriptscriptstyle\text{\tiny-}k}
\def\subnk{_{\scriptscriptstyle \nk}}
\def\subk{_{\scriptscriptstyle k}}
\def\subnbk{_{\scriptscriptstyle [\text{\tiny -}k]}}
\def\subbk{_{\scriptscriptstyle [k]}}
\def\subnbkkp{_{\scriptscriptstyle[\nk,\nk\text{\tiny '}]}}

\def\Dsc{\mathcal{D}}

\def\Lsc{\mathcal{L}}
\def\bD{\mathbf{D}}

\def\Lsc{\mathcal{L}}

\def\Lschat{\widehat{\Lsc}}
\def\argmindum{\mathop{\mbox{argmin}}}
\def\argmin#1{\argmindum_{#1}}

\definecolor{purple}{rgb}{0.84, 0.17, 0.89}

\def\nminv{n_m^{-1}}
\def\suminm{\sum_{i=1}^{n_m}}
\def\Ninv{N^{-1}}
\def\Isc{\mathcal{I}}
\def\Pscr{\mathscr{P}}
\def\Dsc{\mathcal{D}}
\def\Bscr{\mathscr{B}}

\def\suponetrans{^{\scriptscriptstyle \sf (1) T}}

\def\supmtrans{^{\scriptscriptstyle \sf {\tiny (}m{\tiny)} T}}

\def\supMtrans{^{\scriptscriptstyle \sf (M) T}}

\def\supone{^{\scriptscriptstyle \sf (1)}}

\def\supm{^{\scriptscriptstyle \sf \text{\tiny(}m\text{\tiny)}}}

\def\supbullet{^{\scriptscriptstyle \sf (\bullet)}}

\def\supM{^{\scriptscriptstyle \sf \text{\tiny(}M\text{\tiny)}}}
\def\supth{^{\scriptscriptstyle \sf th}}

\def\bX{\mathbf{X}}

\def\Tbb{\mathbb{T}}

\def\Phibar{\bar{\Phi}}

\def\bzero{\mathbf{0}}
\def\Obb{\mathbb{O}}

\newtheorem{definition}{Definition}
\newtheorem{lemma}{Lemma}

\newtheorem{assumption}{Assumption}

\newtheorem{theorem}{Theorem}
\newtheorem{remark}{Remark}

\newcommand{\beq}{\begin{equation}}
\newcommand{\eeq}{\end{equation}}
\newcommand{\beas}{\begin{eqnarray*}}
\newcommand{\eeas}{\end{eqnarray*}}
\newcommand{\bea}{\begin{eqnarray}}
\newcommand{\eea}{\end{eqnarray}}

\newcommand{\bet}{\begin{theorem}}
\newcommand{\eet}{\end{theorem}} 
\newcommand{\bel}{\begin{lemma}}
\newcommand{\eel}{\end{lemma}}
\newcommand{\bep}{\begin{proposition}}
\newcommand{\eep}{\end{proposition}}
\newcommand{\bed}{\begin{definition}}
\newcommand{\eed}{\end{definition}}
\newcommand{\bec}{\begin{corollary}}
\newcommand{\eec}{\end{corollary}}
\newcommand{\bex}{\begin{example}}
\newcommand{\eex}{\end{example}}

\usepackage{color}

\newcommand{\bei}{\begin{itemize}}
\newcommand{\eei}{\end{itemize}}
\newcommand{\ben}{\begin{enumerate}}
\newcommand{\een}{\end{enumerate}}
\def\0{\boldsymbol{0}}
\def\u{\boldsymbol{u}}

\def\x{\mathbf{x}}

\def\X{\boldsymbol{X}}

\def\S{\boldsymbol{\Sigma}}

\def\a{\boldsymbol{a}}

\def\x{\boldsymbol{x}}
\def\bxi{\boldsymbol{\xi}}
\def\Bcal{\mathcal{B}}
\def\Ubb{\mathbb{U}}

\newcommand{\RR}{\mathbb{R}}

\newbox\TempBox \newbox\TempBoxA

\def\pr{\textsf{P}} 

\def\ep{\textsf{E}} 

\def\trans{^{\scriptscriptstyle \sf T}}

\def\text#1{\mbox{\sf #1}}

\def\bX{\mathbf{X}}
\def\half{\frac{1}{2}}

\def\bbeta{\boldsymbol{\beta}}
\def\boeta{\boldsymbol{\boeta}}

\def\bbetahat{\widehat{\bbeta}}

\def\bep{\boldsymbol{\epsilon}}

\def\half{\frac{1}{2}}
\def\Ssc{\mathcal{S}}
\def\Sschat{\widehat{\Ssc}}
\def\cH{\mathcal{H}}

\def\Hbbhat{\widehat{\Hbb}}
\def\Hbb{\mathbb{H}}

\def\bbetatilde{\widetilde{\bbeta}}
\def\betatilde{\widetilde{\beta}}

\def\Pscrhat{\widehat{\Pscr}}

\def\Hbbtilde{\widetilde{\Hbb}}
\def\u{\boldsymbol{u}}
\def\uhat{\widehat{\u}}
\def\e{\boldsymbol{e}}

\def\betabreve{\breve{\beta}}

\def\sigmahat{\widehat{\sigma}}
\def\S{\boldsymbol{\Sigma}}

\def\zetabreve{\breve{\zeta}}
\def\Hsc{\mathcal{H}}
\def\Jbb{\mathbb{J}}
\def\Jbbtilde{\widetilde{\Jbb}}
\def\Asc{\mathcal{A}}

\usepackage{mathtools}

\newcommand\blfootnote[1]{%
  \begingroup
  \renewcommand\thefootnote{}\footnote{#1}%
  \addtocounter{footnote}{-1}%
  \endgroup
}

\begin{document}

\setstretch{1.2}
\def\spacingset#1{\renewcommand{\baselinestretch}%
{#1}\small\normalsize} \spacingset{1}

\title{Integrative High Dimensional Multiple Testing with Heterogeneity under Data Sharing Constraints}

\author{Molei Liu$^{1*}, \blfootnote{Joint first author.}$ Yin Xia$^{2*}$, \ Kelly Cho$^{3}$ \ and \ Tianxi Cai$^{1}$   \ }

\date{}

\maketitle

\footnotetext[1]{Department of Biostatistics, Harvard Chan School of Public Health.}

\footnotetext[2]{Department of Statistics, School of Management, Fudan University}

\footnotetext[3]{Massachusetts Veterans Epidemiology Research and Information Center, US Department of Veteran Affairs}

\begin{abstract}


Identifying informative predictors in a high dimensional regression model is a critical step for association analysis and predictive modeling. Signal detection in the high dimensional setting often fails due to the limited sample size. One approach to improve power is through meta-analyzing multiple studies on the same scientific question. However, integrative analysis of high dimensional data from multiple studies is challenging  in the presence of between study heterogeneity. The challenge is even more pronounced with additional data sharing constraints under which only summary data but not individual level data can be shared across different sites.  In this paper, we propose a novel  data shielding integrative large-scale testing approach to signal detection by allowing between study heterogeneity and not requiring sharing of individual level data. Assuming the underlying high dimensional regression models of the data differ across studies yet share  similar support, 
the proposed method incorporates proper integrative estimation and debiasing procedures to construct test statistics for the overall effects of specific covariates. We also develop a multiple testing procedure to identify significant effects while controlling the false discovery rate and false discovery proportion. Theoretical comparisons of the new testing procedure with the ideal individual--level meta--analysis approach and other distributed inference methods are investigated.  
Simulation studies demonstrate that the proposed testing procedure performs well in both controlling false discovery and attaining power. The new method is applied to a real example on detecting interaction effect of the genetic variants for statins and obesity on the risk for type II diabetes.

\end{abstract}

\noindent {\bf Keywords:} Debiasing; Distributed learning;  False discovery rate; High dimensional inference; Integrative analysis; Multiple testing.



\newpage

\section{Introduction}\label{sec:intro}
\subsection{Background}

High throughput technologies such as genetic sequencing and natural language processing have led to an increasing number and types of predictors available to assist in predictive modeling. A critical step in developing accurate and robust prediction model is to differentiate true signals from noise. A wide range of high dimensional inference procedures have been developed in recent years to achieve variable selection, hypothesis testing and interval estimation \citep[e.g.]{van2014asymptotically,javanmard2014confidence,zhang2014confidence,chernozhukov2018double}. However, regardless of the procedure, drawing precise high dimensional inference is often infeasible in practical settings where the available sample size is too small relative to the number of predictors. One approach to improve the precision and boost power is through meta-analyzing multiple studies that address underlying the same scientific problem. This approach has been widely adopted in practice in many scientific fields, including clinical trials, education, policy evaluation,  ecology, and genomics \cite[e.g.]{dersimonian1996meta,allen2002comparing,card2010active,stewart2010meta,panagiotou2013power}, as tools for evidence based decision making. Meta-analysis is particularly valuable in the high dimensional setting. For example, meta-analysis of high dimensional genomic data from multiple studies has uncovered new disease susceptibility loci for a broad range of diseases including crohn's disease, colorectal cancer, childhood obesity and type II diabetes \citep[e.g.]{study2008meta,bradfield2012genome,franke2010genome,zeggini2008meta}.

Integrative analysis of high dimensional data, however, is highly challenging especially with biomedical studies for several reasons. First, between study heterogeneity arises frequently due to the difference in patient population and data acquisition. Second, due to privacy and legal constraints, individual level data often cannot be shared across study sites but rather only summary statistics can be passed between researchers. For example, patient level genetic data linked with clinical variables extracted from electronic health records of several hospitals are not allowed to leave the firewall of each hospital. In addition to high dimensionality, attentions to both heterogeneity and data sharing constraints are needed to perform meta-analysis of multiple electronic health records linked genomic studies. 

The aforementioned data sharing mechanism is referred to as DataSHIELD  (Data aggregation through anonymous Summary-statistics from Harmonised Individual levEL Databases) in \cite{wolfson2010datashield}, which has been widely accepted as a useful strategy to protect patient privacy  \citep{jones2012datashield,doiron2013data}.  Several statistical approaches to integrative analysis under the DataSHILED framework have been developed for low dimensional settings \citep[e.g.]{gaye2014datashield,zoller2018distributed,tong2020robust}. In the absence of cross-site heterogeneity, distributed high dimensional estimation and inference procedures have also been  developed recently that can facilitate DataSHIELD constraints  \citep[e.g.]{lee2017communication,battey2018distributed,jordan2018communication}. Recently, \cite{cai2019privacy} proposed an integrative high dimensional sparse regression approach that accounts for heterogeneity. However, their method is limited to parameter estimation and variable selection. To the best of our knowledge, no hypothesis testing procedures currently exist to enable identification of significant predictors with false discovery error control under the setting of interest. In this paper, we propose a data shielding integrative large-scale testing procedure to fill this gap.

\subsection{Problem statement}\label{sec:intro:state}
Suppose there are $M$ independent studies and the $m$th study contains observations on an outcome $Y\supm$ and a $p$-dimensional covariate vector $\X\supm$, where $Y\supm$ can be binary or continuous and without loss of generality we assume that $\X\supm$ contains 1 as its first element. Specifically, data from the $m$th study consist of $n_m$ independent and identically distributed random vectors, $\Dsc\supm  = \{\bD\supm_i  = (Y_i\supm, \bX_i\supmtrans)\trans, i = 1, ..., n_m\}$. Let $N=\sum_{m=1}^Mn_m$ and $n = N/M$. We assume a generalized linear model
${\rm E}(Y\supm \mid \X\supm) = g(\bbeta_0\supmtrans\X\supm)$ and the true model parameter $\bbeta_0\supm$ is the minimizer of the population loss function: 
$$
\bbeta_0\supm= \argmin{\bbeta\supm\in\mathbb{R}^p}{\Lsc_m(\bbeta\supm)},~\mbox{where}~\Lsc_m(\bbeta\supm) = \ep\{f(\bX_i\supmtrans\bbeta\supm,Y_i\supm)\}, \quad {f(x,y) = \phi(x) -yx}, 
$$
where $\dot{\phi}(x)  \equiv d \phi(x) /d x= g(x)$. When $\phi(x)=\log(1+e^x)$, this  corresponds to logistic model  if $Y$ is binary and a quasi-binomial model if $Y \in [0,1]$ is a continuous probability
metric sometimes generated from an electronic health records probabilistic phenotyping algorithm. We may let $\phi(x) = x^2/2$ to correspond the standard least square for continuous $Y$. Under the DataSHIELD constraints, the individual--level data $\Dsc\supm$ is stored at the $m\supth$ data computer and only summary statistics are allowed to transfer from the distributed data computers to the analysis computer as the central node. 

Our goal is to develop procedures under the DataSHIELD constraints for testing
\beq\label{test}
H_{0,j}: \bbeta_{0,j}\equiv (\beta^{\scriptscriptstyle \sf(1)}_{0,j}, \ldots, \beta\supM_{0,j})\trans = \bzero \mbox{ v.s. } H_{a,j}: \bbeta_{0,j} \ne \bzero
\eeq
simultaneously for $j\in \cH$ to identify $\cH_1 = \{j \in \Hsc: \bbeta_{0,j}\ \neq \bzero\}$, while controlling the false discovery rate and false discovery proportion, where $\cH \subseteq \{2,\ldots,p\}$ is a user-specified subset with $|\cH|=q \asymp p$ and $|\Asc|$ denotes the size of any set $\Asc$. Here $\bbeta_{0,j}= \bzero$ indicates that $X_j$ is independent of $Y$ given all remaining covariates.
To ensure effective integrative analysis, we assume that $\bbeta_0\supone, ..., \bbeta_0\supM$ are sparse and share similar support. Specifically, we assume that $|\Ssc_0| \ll p$ and $s\supm\asymp s$ for $m=1,2,\ldots,M$, where $\Ssc_0 = \{j=2,...,p: \bbeta_{0,j}\supm \ne \bzero\} = \cup_{m =1}^M \Ssc\supm$, $\Ssc\supm = \{j=2,...,p: \beta_{0,j}\supm \ne 0\}$, $s\supm = |\Ssc\supm|$, and $s = |\Ssc_0|$.

\subsection{Our contribution and the related work}
We propose in this paper a novel data shielding integrative large-scale testing procedure with false discovery rate and false discovery proportion control for the simultaneous inference problem \eqref{test}.
The proposed testing procedure consists of three major steps: (I) derive an integrative estimator on the analysis computer using locally obtained summary statistics from the data computers and send the estimator back to the local sites; (II) construct group effect test statistic for each covariate through an integrative debiasing method; and (III) develop an error rate controlled multiple testing procedure based on the group effect statistics. 

The integrative estimation approach in the first step is closely related to the group inference methods in the literature.
Denote by $\bbeta_j=(\beta\supone_j,...,\beta\supM_j)\trans$, $\bbeta\supbullet=(\bbeta\suponetrans,\ldots,\bbeta\supMtrans)\trans$,
\[
\Lschat\supm(\bbeta\supm)=\nminv\suminm f(\bbeta\supmtrans\bX_i\supm,Y\supm_i)\quad\mbox{and}\quad\Lschat\supbullet(\bbeta\supbullet)=\Ninv \sum_{m=1}^Mn_m\Lschat_m(\bbeta\supm).
\]
Literature in group LASSO and multi-task learning \citep[e.g.]{huang2010benefit,lounici2011oracle} established that, under the setting $s\supm\asymp s$ as introduced in Section \ref{sec:intro:state}, the group LASSO estimator with tuning parameter $\lambda$:
$
\argmin{\bbeta\supbullet}\Lschat\supbullet(\bbeta\supbullet)+\lambda\sum_{j=2}^p\|\bbeta_j\|_2, 
$
benefits from the group structure and attains the optimal rate of convergence. In this paper, we adopt the same structured group LASSO penalty for integrative estimation, but under the data sharing constraints. Recently, \cite{mitra2016benefit} proposed a group structured debiasing approach under the integrative analysis setting, where they restricted their analysis to linear models and required strong group sparsity assumptions on the covariates of the distributed datasets. 
In contrast,  our method accommodates generalized linear models and has more relaxed assumption on the generation mechanism of the design matrix.



The second step of our method, i.e., the construction of the test statistics for each of the hypotheses, relies on the group debiasing of the above integrative estimation. For debiasing of generalized linear models, nodewise LASSO regression was employed in the earlier work \citep[e.g]{van2014asymptotically,jankova2016confidence}, while Dantzig selector type approach was proposed more recently \citep[e.g]{belloni2018high,caner2018high}.
We develop in this article a cross--fitted group Dantzig selector type debiasing method, which requires weaker inverse Hessian assumptions than the aforementioned approaches.
In addition, the proposed debiasing step achieves proper bias rate under the same model sparsity assumptions as the ideal individual--level meta--analysis method. Compared with the one--shot distributed inference approaches \citep{tang2016method,lee2017communication,battey2018distributed}, the proposed method additionally considers model heterogeneity and group inference; it further reduces the bias rate by sending the integrative estimator to the data computers to derive updated summary statistics, which in turn benefits the subsequent multiple testing procedure. See Section \ref{sec:thm:com} for detailed comparisons. 
As the last step, simultaneous inference with theoretical error rates control was performed based on the group effect statistics. The test statistics are shown to be asymptotically chi-square distributed under the null, and the proposed multiple testing procedure asymptotically controls both the false discovery rate and false discovery proportion at the pre-specified level.
Multiple testing for high-dimensional regression models has been recently popularly studied in the literature \citep[e.g]{liu2014hypo,xia2018two,xia2018joint}. The simultaneous inference approach we establish in this article is considerably different from the aforementioned papers in the following aspects. First, we consider a more general M-estimation setting which can accommodate different types of outcomes, while the earlier works mostly focus on the linear models. Second, in contrast to the joint inference considered in \cite{xia2018joint} where they assumed the covariate homogeneity and common support of the regression coefficients, we allow the heterogeneity in both the covariates and the coefficients. Third, the DataSHIELD framework only allows the communication of summary statistics, and thus the existing testing approaches developed for individual-level data are no longer suitable to the current setup. Last, the novel integrative testing procedure brings new technical difficulties on the theoretical error rates studies under the complicated dependence structures. Hence, our proposal makes a useful addition to the general toolbox of simultaneous regression inference.

We also demonstrate via numerical experiments that the proposed testing procedure attains good power while maintaining error rate control. In addition, our new approach outperforms existing distributed inference methods and enjoys similar performance as the ideal individual--level meta--analysis approach. The proposed method is further applied to a real example. 





\section{Data shielding integrative large-scale testing procedure}\label{sec:method}
\subsection{Notation}\label{sec:method:not}
Throughout, for any integer $d$, any vector $\x=(x_1,x_2,\ldots,x_d)\trans\in \RR^d$, and any set $\Ssc = \{j_1, \ldots, j_k\}\subseteq [d] \equiv \{1, \ldots, d\}$,  denote by $\x_{\Ssc} = [x_{j_1}, \ldots, x_{j_k}]'$, $\x_{\text{-}j}$ the vector with its $j\supth$ entry removed from  $\x$, $\|\x\|_q$ the $\ell_q$ norm of $\x$ and $\|\x\|_{\infty}=\max_{j\in[d]}|x_j|$. 
For any $d$-dimensional vectors $\{\a\supm=(a\supm_1,\ldots,a\supm_d)\trans, m \in [M]\}$ and $\Ssc \subseteq [d]$, let $\a\supbullet=(\a\suponetrans,\ldots,\a\supMtrans)\trans$, $\a_{\Ssc}\supbullet=(\a\suponetrans_{\Ssc},\ldots,\a\supMtrans_{\Ssc})\trans$, $\a_j = (a\supone_j, \ldots, a\supM_j)\trans$,
and $\a_{\text{-}j}\supbullet=(\a\suponetrans_{\text{-}j},\ldots,\a\supMtrans_{\text{-}j})\trans$. Let $\e_j$ be the unit vector with $j\supth$ element being $1$ and remaining elements being $0$ and $\e_j\supbullet=(\e_j\trans,\ldots,\e_j\trans)\trans$. 
Denote  by $\|\a\supbullet\|_{2,1}=\sum_{j=1}^d\|\a_j\|_2$ and $\|\a\supbullet\|_{2,\infty}=\max_{j\in[d]}\|\a_j\|_2$ the $\ell_2/\ell_1$ and $\ell_2/\ell_{\infty}$ norm of $\a\supbullet$ respectively. For any K-fold partition of $[n_m]$, denoted by $\{\Isc_k\supm, k\in[K]\}$, let 
 $\Isc\subnk\supm=[n_m]\setminus\Isc_k\supm$, $\Isc_k\supbullet=\{\Isc_k\supm:m\in[M]\}$,  $\Isc\subnk\supbullet=\{\Isc\subnk\supm:m\in[M]\}$. For any index set $\Isc\supbullet = \{\Isc\supm \subseteq[n_m], m \in [M]\}$, $\Dsc_{\Isc\supm}\supm = \{\bD_i\supm:i\in\Isc\supm\}$, and $\Dsc_{\Isc\supbullet}\supbullet=\{\Dsc_{\Isc\supm}\supm:m\in[M]\}$. 
  Let $\ddot{\phi}(\theta) = d^2 \phi(\theta)/ d\theta^2 \ge 0$. Denote by $\bbeta_{0,j}$ and $\bbeta_0\supbullet$ the true values of $\bbeta_j$ and $\bbeta\supbullet$ respectively.  For any $\Isc\supbullet$ and $\bbeta\supbullet$, define the sample measure operators $\Pscrhat_{\Isc\supm}\eta_{\bbeta\supm}=|\Isc\supm|^{-1}\sum_{i\in\Isc\supm}\eta_{\bbeta\supm}(\bD_i\supm)$ and $\Pscrhat_{\Isc\supbullet}\eta_{\bbeta\supbullet}=|\Isc\supbullet|^{-1}\sum_{m=1}^M\sum_{i\in\Isc\supm}\eta_{\bbeta\supm}(\bD_i\supm)$, and the population measure operator  $\Pscr\supm \eta_{\bbeta\supm}={\sf E}\eta_{\bbeta\supm}(\bD\supm_i)$, 
for all integrable functions $\eta_{\bbeta\supbullet} = \{\eta_{\bbeta\supm}, m \in [M]\}$ parameterized by  $\bbeta\supbullet$ or $\bbeta\supm$.

For any given $\bbeta\supm$, we define $\theta\supm_i=\bX\supmtrans_i\bbeta\supm$, $\theta\supm_{0,i}=\bX\supmtrans_i\bbeta_0\supm$, and the residual $\epsilon_i\supm:=Y_i\supm-\dot{\phi}(\theta\supm_{0,i})$. Similar to \cite{cai2019differential} and \cite{ma2020global}, given coefficient $\bbeta\supm$, we can express $Y_i\supm\sim\bX_i\supm$ in an approximately linear form:
\[
Y_i\supm-\dot{\phi}(\theta\supm_i)+\ddot{\phi}(\theta\supm_i)\theta\supm_i=\ddot{\phi}(\theta\supm_i)\bX_i\supmtrans\bbeta\supm_0+\epsilon_i\supm+R\supm_i(\theta\supm_i),
\]
where $R\supm_i(\theta\supm_i)$ is the reminder term and $R\supm_i(\theta\supm_{0,i})=0$. For a given observation set $\bD$ and coefficient $\bbeta$, we let $\theta=\bX\trans\bbeta$, $Y_{\bbeta}=\ddot{\phi}^{-\frac{1}{2}}(\theta)\left\{Y-\dot{\phi}(\theta)+\ddot{\phi}(\theta)\theta\right\}$, $\bX_{\bbeta}=\ddot{\phi}^{\frac{1}{2}}(\theta)\bX$. 
Note that for logistic model, we have ${\rm Var}(Y_{\bbeta}|\bX_{\bbeta})=1$ and $\bX_{\bbeta}$ and $Y_{\bbeta}$ can be viewed as the covariates and responses adjusted for the heteroscedasticity of the residuals.



\subsection{Outline of the proposed testing procedure}
We first outline in this section the data shielding integrative large-scale testing procedure in Algorithm \ref{alg:meta} and then study the details of each key step later in Sections \ref{sec:method:int} to \ref{sec:method:mult}.
The procedure involves partitioning of $\Dsc\supm$ into $K$ folds $\{\Isc_k\supm:k \in [K]\}$ for $m\in[M]$, where without loss of generality we let $K\ge 2$ be an even number. With a slight abuse of notation, we write $\Dsc\subbk \supm = \Dsc_{\Isc\subk\supm}\supm$,  $\Dsc\subbk\supbullet =\Dsc_{\Isc\subk\supbullet}\supbullet$, $\Dsc\subnbk \supm = \Dsc_{\Isc\subnk\supm}\supm$, and $\Dsc\subnbk\supbullet =\Dsc_{\Isc\subnk\supbullet}\supbullet$.

\begin{algorithm}[htbp]
\caption{\label{alg:meta} Data Shielding Integrative Large-scale Testing Algorithm.}
{Input: }$\Dsc\supm$ at the $m\supth$ data computer for $m\in[M]$.
\begin{enumerate}[Step 1.]
\item \label{alg1:step1} For each $k \in [K]$, fit {\bf integrative sparse regression under DataSHIELD} with $\Dsc\subnbk\supbullet$: 
\begin{enumerate}[(a)]
\item At the $m\supth$ data computer, construct cross-fitted summary statistics based on local LASSO estimator, and send them to the analysis computer; 
\item At the analysis computer, obtain the integrative estimator $\bbetatilde\subnbk\supbullet$, and send them back to each data computer.
\end{enumerate}

\item \label{alg1:step2} {\bf Obtain debiased group test statistics}:
\begin{enumerate}[(a)]
\item For each $k$, at the $m\supth$ data computer, obtain the updated summary statistics based on $\bbetatilde\supbullet\subnbk$ and $\Dsc\subbk\supm$, and send them to the analysis computer; 
\item At the analysis computer, construct a set of cross-fitted debiased group estimators $\{\zetabreve_j, j\in \cH\}$.
\end{enumerate}

\item \label{alg1:step3} Construct a multiple testing procedure based on the test statistics from Step \ref{alg1:step2}.
\end{enumerate}

\end{algorithm}

\subsection{Step \ref{alg1:step1}: Integrative sparse regression}\label{sec:method:int}
As a first step, we fit integrative sparse regression under DataSHIELD with $\Dsc\subnbk\supbullet$ following similar strategies as given in \cite{cai2019privacy}. To carry out Step 1(a)  of Algorithm \ref{alg:meta}, we split the index set $\Isc\subnk\supm$ into $K'$ folds $\Isc_{\text{\tiny-}k,1}\supm,\ldots,\Isc_{\text{\tiny-}k,K'}\supm$. For $k \in [K]$ and $k' \in [K']$, we construct local LASSO estimator with tuning parameter $\lambda\supm$: 
$
\bbetahat\supm\subnbkkp=\argmin{\bbeta\supm\in\mathbb{R}^p}\Pscrhat_{\Isc\subnk\supm\setminus\Isc_{\text{\tiny-}k,k'}\supm}f(\bX\trans\bbeta\supm,Y)+\lambda\supm\|\bbeta\supm_{\text{-}1}\|_1.
$
With $\Dsc\subnbk\supm$, we then derive summary data  $\Ssc\subnbk\supm = \{|\Isc\subnk\supm|, \widehat{\bxi}\supm\subnbk, \Hbbhat\supm\subnbk\}$, 
where 
\begin{equation}
\widehat{\bxi}\supm\subnbk={K'}^{-1}\sum_{k'=1}^{K'}\Pscrhat_{\Isc_{\text{\tiny-}k,k'}\supm}\bX_{\bbetahat\supm\subnbkkp}Y_{\bbetahat\supm\subnbkkp},\quad
\Hbbhat\supm\subnbk={K'}^{-1}\sum_{k'=1}^{K'}\Pscrhat_{\Isc_{\text{\tiny-}k,k'}\supm}\bX_{\bbetahat\supm\subnbkkp}\bX_{\bbetahat\supm\subnbkkp}\trans.
\label{equ:sum}
\end{equation}
In Step \ref{alg1:step1}(b) of Algorithm \ref{alg:meta}, for $k \in [K]$, we aggregate the $M$ sets of summary data $\{\Ssc\subnbk\supm, m\in[M]\}$ at the central analysis computer and solve a regularized quasi--likelihood problem to obtain the integrative estimator with tuning parameter $\lambda$:
\begin{equation}
\bbetatilde\supbullet\subnbk=\argmin{\bbeta\supbullet}|\Isc\supbullet\subnk|^{-1}\sum_{m=1}^M|\Isc\subnk\supm|\left(\bbeta\supmtrans\Hbbhat\supm\subnbk\bbeta\supm-2\bbeta\supmtrans\widehat{\bxi}\supm\subnbk\right)+\lambda\|\bbeta_{\text{-}1}\supbullet\|_{2,1}.
\label{equ:adele}
\end{equation}
These $K$ sets of estimators, $\{\bbetatilde\supbullet\subnbk, k \in [K]\}$, are then sent back to the data computers. 
The summary statistics introduced in (\ref{equ:sum}) can be viewed as the covariance terms of $\Dsc\subnbk\supm$ with the local LASSO estimator plugged-in to adjust for the heteroscedasticity of the residuals. Cross--fitting is used to remove the dependence of the observed data and the fitted outcomes - a strategy frequently employed in high dimensional inference literatures \citep{chernozhukov2016double,chernozhukov2018double}. {As in \cite{cai2019privacy}, the integrative procedure can also be viewed in such a way that $\bbeta\supmtrans\Hbbhat\supm_{[\text{-}k]}\bbeta\supm-2\bbeta\supmtrans\widehat{\bxi}\supm_{[\text{-}k]}$ provides a second order one--step approximation to the individual--level data loss function $2\Pscrhat_{\Isc\subnk\supm}f(\bX\trans\bbeta\supm,Y)$ initializing with the local LASSO estimators.} {In contrast to \cite{cai2019privacy}, we introduce cross--fitting procedure at each local data computer to reduce fitting bias and this in turn relaxes their uniformly bounded assumption on $\bX_i\supmtrans\bbeta\supm$ for each $i$ and $m$, i.e., Condition 4(i) of \cite{cai2019privacy}.}


\subsection{Step \ref{alg1:step2}: Debiased group test statistics}\label{sec:method:de}

We next derive group effect test statistics in Step \ref{alg1:step2} by constructing debiased estimators for $\bbeta_0\supbullet$ and estimating their variances. In Step \ref{alg1:step2}(a), we construct updated summary statistics 
$$
\widetilde{\bxi}\supm\subbk=\Pscrhat_{\Isc_k\supm}\bX_{\bbetatilde\supm_{[\text{\tiny-}k]}}Y_{\bbetatilde\supm_{[\text{\tiny-}k]}},~~\Hbbtilde\supm\subbk=\Pscrhat_{\Isc_k\supm}\bX_{\bbetatilde\supm_{[\text{\tiny-}k]}}\bX_{\bbetatilde\supm_{[\text{\tiny-}k]}}\trans~~\mbox{and}~~\Jbbtilde\subbk\supm=\Pscrhat_{\Isc_k\supm}\bX\bX\trans\left\{Y-\dot\phi(\bX\trans\bbetatilde\supm_{[\text{-}k]})\right\}^2
$$ 
at the $m$th data computer, for $k \in [K]$. These $mK$ sets of summary statistics are then sent to the analysis computer in Step \ref{alg1:step2}(b) to be aggregated and debiased. Specifically, 
for each $j\in \cH$ and $k \in [K]$, we solve the group dantzig selector type optimization problem:
\begin{equation}
\uhat_{j,[k]}\supbullet=\argmin{\u\supbullet} \max_{m\in [M]}\|\u\supm\|_1\quad\mbox{s.t.}\quad\|\Hbbtilde\supbullet_{[k]}\u\supbullet-\e_j\supbullet\|_{2,\infty}\leq\tau,
\label{equ:deb}
\end{equation}
to obtain a vector of projection directions for some tuning parameter $\tau$, where $\Hbbtilde\supbullet_{[k]}={\sf diag}\{\Hbbtilde\supone_{[k]},\ldots,\Hbbtilde\supM_{[k]}\}$. Combining across the $K$ splits, we 
construct the cross--fitted group debiased estimator for $\beta_j\supm$ by $\betabreve_{j}\supm=K^{-1}\sum_{k=1}^K  \left\{\betatilde\supm_{j,[\text{-}k]}+\uhat_{j,[k]}\supmtrans(\widetilde{\bxi}\supm_{[k]}-\Hbbtilde\supm_{[k]}\bbetatilde\supm_{[\text{-}k]})\right\}$.

In Section \ref{sec:thm:deb}, we show that the distribution of $n_m^{1/2}(\betabreve_j\supm-\beta_{0,j})$ is approximately normal with mean 0 and variance  $(\sigma_{0,j}\supm)^2$, which can be estimated by $(\sigmahat_j\supm)^2=K^{-1}\sum_{k=1}^K\uhat_{j,[k]}\supmtrans\Jbbtilde_{[k]}\supm\uhat_{j,[k]}\supm$. Finally, we test for the group effect of the $j$-th covariate across $M$ studies based on the standardized sum of square type statistics $\zetabreve_j = \sum_{m=1}^Mn_m\{\betabreve_{j}\supm/\sigmahat\supm_j\}^2$ for $j\in \cH$.

We show in Section  \ref{sec:thm:deb}  that, under mild regularity assumptions, $\zetabreve_j$ is asymptotically chi-square distributed with degree of freedom $M$ under the null. This result is crucial to ensure the error rate control for the downstream multiple testing procedure. 

\def\Nj{{\mathcal N}_j}
\def\Fbb{\mathbb{F}}

\subsection{Step \ref{alg1:step3}: Multiple testing}\label{sec:method:mult}
To construct an error rate controlled multiple testing procedure for
\[
H_{0,j}: \bbeta_{0,j} = \textbf{0} \mbox{ versus } H_{a,j}: \bbeta_{0,j} \neq \textbf{0}, \quad\mbox{  } j \in \cH \subseteq \{2,\ldots,p\},
\]
we first take a normal quantile transformation of $\zetabreve_j$, namely
$
\Nj = \bar\Phi^{-1}\left\{\bar\Fbb_{\chi^2_M}(\zetabreve_j)/2\right\},
$
where $\Phi$ is the standard normal cumulative distribution function, $\bar\Phi = 1-\Phi$, and $\bar\Fbb_{\chi^2_M}(\cdot)$ is the survival function of $\chi^2_M$. Based on the asymptotic $\chi^2_M$ distribution of $\zetabreve_j$ as will be shown in Theorem \ref{thm:1}, we present in the proof of Theorem \ref{FDR1} that $\Nj$ asymptotically has the same distribution as the absolute value of a standard normal random variable. Thus, to test a single hypothesis of $H_{0,j}: \bbeta_{0,j} = \textbf{0}$, we reject the the null at nominal level $\alpha>0$ whenever $\Psi_{\alpha,j}=1$, where
$
\Psi_{\alpha,j}=I\left\{\Nj\geq \bar\Phi^{-1}(\alpha/2)\right\}.
$

However, for simultaneous inference across $q$ hypotheses $\{H_{0,j}, j \in \cH\}$, we shall further adjust the multiplicity of the tests as follows.
For any threshold level $t$, let 
$R_{0}(t) = \sum_{j\in \mathcal{H}_0}I(\Nj\geq t)$ and $R(t)= \sum_{j\in \cH}I(\Nj\geq t)$ respectively denote the total number of false positives and the total number of rejections associated with $t$, where $\cH_0 = \{j \in \Hsc: \bbeta_{0,j}\ = \bzero\}$.
Then the false discovery proportion and false discovery rate for a given $t$ are respectively defined as 
\[
\text{FDP}(t)=\frac{R_{0}(t)}{R(t)\vee 1} \quad \mbox{and}\quad \text{FDR}(t)=\ep\{\text{FDP}(t)\}.
\]
The smallest $t$ such that $\text{FDP}(t)\leq \alpha$, namely $t_0=\inf\left\{0\leq t\leq (2\log q)^{1/2}: \; \text{FDP}(t)\leq \alpha\right\}$ would be a desirable threshold since it maximizes the power under the false discovery proportion control. However, since the null set is unknown, we 
estimate $R_{0}(t)$ by $2\bar\Phi(t)|\mathcal{H}_0|$ and conservatively estimate $|\mathcal{H}_0|$ by $q$ because of the model sparsity. 
We next calculate 
\beq\label{t_hat}
\hat{t}=\inf\left\{0\leq t\leq t_q: \;  \frac{2q\bar\Phi(t)}{R(t)\vee 1}\leq \alpha\right\} \quad \mbox{where}\quad t_q = (2\log q-2\log\log q)^{\frac{1}{2}}
\eeq
to approximate the ideal threshold $t_0$.
If (\ref{t_hat}) does not exist, we set $\hat{t}=(2\log q)^{1/2}$. Finally, we obtain the rejection set $\{j: \Nj\geq \hat{t}, j\in \cH\}$ as the output of Algorithm \ref{alg:meta}. The theoretical analysis of the asymptotic error rates control of the proposed multiple testing procedure will be studied in Section \ref{sec:thm:fdr}.

\def\blambda{\boldsymbol{\lambda}}
\def\bdeta{\boldsymbol{\eta}}
\def\dhat{\widehat{d}}
\def\Rhat{\widehat{R}}

\subsection{Tuning parameter selection}\label{sec:method:tune}
In this section, we detail data-driven procedures for selecting the tuning parameters $\bdeta = \{\blambda\supbullet = (\lambda\supone,\ldots,\lambda\supM)\trans, \lambda, \tau\}$.  Since our primary goal is to perform simultaneous testing, we follow a similar strategy as that of \cite{xia2018joint} and select tuning parameters to minimize a $\ell_2$ distance between $\Rhat_0(t)/\{2|\cH_0|\bar\Phi(t)\}$ and its expected value of 1, where $\Rhat_0(t)$ is an estimate of $R_0(t)$ from the testing procedure.  However, unlike \cite{xia2018joint}, it is not feasible to tune $\bdeta$ simultaneously due to DataSHILED constraint. We instead tune $\blambda\supbullet$, $\lambda$ and $\tau$ sequentially as detailed below. Furthermore, based on the theoretical analyses of the optimal rates for $\bdeta$ given in Section \ref{sec:thm}, we select $\bdeta$ within a set of candidate values that are of the same order as their respective optimal rates. 

First for $\blambda\supbullet$ in Algorithm \ref{alg:meta}, we tune $\lambda\supm$ via cross validation within the $m$th data computer. Second, to select $\lambda$ for the integrative estimation in (\ref{equ:adele}),  we minimizes an approximated generalized information criterion that only involve derived data from $M$ studies. Specifically, we choose $\lambda$ as the minimizer of
$
{\rm GIC}\left(\lambda,\bbetatilde\supbullet_{[\text{-}k],\lambda}\right)={\rm Dev}\left(\bbetatilde\supbullet_{[\text{-}k],\lambda}\right)+\gamma\mbox{DF}\left(\lambda,\bbetatilde\supbullet_{[\text{-}k],\lambda}\right),
$
where $\gamma$ is some pre-specified scaling parameter, $\bbetatilde\supm_{[\text{-}k],\lambda}$ is the estimator obtained with $\lambda
$, 
\begin{align*}
{\rm Dev}\left(\bbeta\supbullet\right) & =|\Isc\subnk|^{-1}\sum_{m=1}^M|\Isc\subnk\supm|\left(\bbeta\supmtrans\Hbbhat\supm_{[\text{-}k]}\bbeta\supm-2\bbeta\supmtrans\widehat{\bxi}\supm_{[\text{-}k]}\right) \quad \mbox{and} \\
\mbox{DF}\left(\lambda,\bbeta\supbullet\right) & =\left[\partial^2_{\Sschat}\left\{{\rm Dev}\left(\bbeta\supbullet\right)+\lambda\|\bbeta\supbullet_{\text{-}1}\|_{2,1}\right\}\right]^{-1}\left[\partial^2_{\Sschat}{\rm Dev}\left(\bbeta\supbullet\right)\right],
\end{align*} 
are respectively the approximated deviance and degree of freedom measures, $\Sschat$ is the set of non-zero elements in $\bbeta\supbullet$ and the operator $\partial^2_{\Sschat}$ denotes the second order partial derivative with respect to $\bbeta\supbullet_{\Sschat}$. Common choices of  $\gamma$ include $2|\Isc\subnk|^{-1}$ (AIC), $|\Isc\subnk|^{-1}\log |\Isc\subnk|$ (BIC), $|\Isc\subnk|^{-1}\log |\Isc\subnk|\log\log p$  \citep[modified BIC]{wang2009shrinkage} and  $2|\Isc\subnk|^{-1}\log |\Isc\subnk|\log p$ \citep[RIC]{foster1994risk}. For numerical studies in Sections \ref{sec:simu} and \ref{sec:real}, we use BIC which appears to perform well across settings. 

\def\smnull{\sf\scriptscriptstyle null}
\def\omegahat{\widehat{\omega}}

At the last step, we tune $\tau$ by minimizing an $\ell_2$ distance between $
\Rhat_{0,\smnull}(t \mid \tau)/\{2q\Phibar(t)\}$ and 1, where $\Rhat_{0,\smnull}(t \mid \tau)$ is an estimate of $R_0(t)$ with a given tuning parameter $\tau$ and we replace $\cH_0$ by $q$ as in  \cite{xia2018joint}. Our construction of $\Rhat_{0,\smnull}(t \mid \tau)$ differs from that of \cite{xia2018joint} in that we estimate $R_0(t)$ under the complete null to better approximate the denominator of $2q\Phibar(t)$.
As detailed in Algorithm \ref{alg:tune}, we construct $\betabreve_{j,\smnull}\supm$ as the difference between the estimator obtained with the  first $K/2$ folds of data and the corresponding estimator obtained using the second $K/2$ folds of data, which is always centered around $0$ rather than $\beta_{0j}\supm$. Since the accuracy of $\Rhat_{0,\smnull}(t \mid \tau)$ for large $t$ is most relevant to the error control, we construct the distance measure $\dhat(\tau)$ in Algorithm \ref{alg:tune} focusing on $t$ around $\Phibar^{-1}[\Phibar\{ (2\log q)^{1/2}\} \iota]$ for some values of $\iota \in (0,1]$.

\def\subhalf{_{\scriptscriptstyle 1/2}}
\begin{algorithm}[htbp]
\caption{\label{alg:tune} Selection of $\tau$ for multiple testing.}
\begin{enumerate}[Step 1.]
\item For any given $\tau$ and each $j\in\cH$, calculate $\zetabreve_{j,{\sf\scriptscriptstyle null}}(\tau)=\sum_{m=1}^Mn_{m}\{\betabreve_{j,{\sf\scriptscriptstyle null}}\supm(\tau)/\sigmahat_{j}\supm\}^2$ with
\begin{align*}
\betabreve_{j,{\sf\scriptscriptstyle null}}\supm(\tau)=&K^{-1}\sum_{k=1}^{K}(-1)^{k >K/2} \left\{\betatilde\supm_{j,[\text{-}k]}+\uhat_{j,[k]}\supmtrans(\tau)\left(\widetilde{\bxi}\supm_{[k]}-\Hbbtilde\supm_{[k]}\bbetatilde\supm_{[\text{-}k]}\right)\right\} , 
\end{align*}
where $\uhat_{j,[k]}\supbullet(\tau)$ is the debiasing projection direction obtained at tuning value $\tau$.

\item Define $\Rhat_{0,\smnull}(t \mid \tau) = \sum_{j\in\cH}I[ \bar\Fbb_{\chi^2_M}\{\zetabreve_{j,{\sf\scriptscriptstyle null}}(\tau)\} \leq 2\bar\Phi(t)]$ and a modified measure
\[
\widehat{d}(\tau)= \int_0^{1} \left[{\Rhat_{0,\smnull}\{\Phibar^{-1}(x)\mid \tau\}}/({2q x})-1\right]^2 d\omegahat(x),
\]
where $\omegahat(x) = H^{-1} \sum_{h=1}^H I(\Phibar\{(2\log q)^{1/2}\} h/H \le x )$ and $H>0$ is some specified constant.
\end{enumerate}
\end{algorithm}



\def\Minv{M^{-1}}
\def\subTconj{_{\widetilde{\Tbb}}}

\section{Theoretical Results}\label{sec:thm}

\subsection{Notation and assumptions}\label{sec:thm:not}
For any semi--positive definite matrix $\mathbb{A}\in\mathbb{R}^{d\times d}$ and $i,j\in[d]$, denote by $\mathbb{A}_{ij}$ the $(i,j)\supth$ element of $\mathbb{A}$ and $\mathbb{A}_{j}$ its $j\supth$ row, $\Lambda_{\min}(\mathbb{A})$ and $\Lambda_{\max}(\mathbb{A})$ the smallest and largest eigenvalue of $\mathbb{A}$. Define the sub-gaussian norm of a random variable $X$ as $\|X\|_{\psi_2}:=\sup_{q\geq1}q^{-1/2}(\ep|X|^q)^{1/q}$
and the sub-gaussian norm of a $d$-dimensional random vector $\X$ as $\|\X\|_{\psi_2}:=\sup_{\x\in\mathbb{S}^{d-1}}\|\x\trans \X\|_{\psi_2}$, where $\mathbb{S}^{d-1}$ is the unit sphere in $\mathbb{R}^{d}$. For $c>0$ and a scalar or vector $\x$, define $\Bcal(\x,c):=\{\x':\|\x'\|_1\leq c\}$ as its $\ell_1$ neighbor with radius $c$. Denote by $\S_0\supm=\Pscr\supm\bX\bX\trans$, $\Hbb_{\bbeta}\supm=\Pscr\supm\bX_{\bbeta}\bX\trans_{\bbeta}$, $\Jbb_{\bbeta}\supm=\Pscr\supm\bX\bX\trans\{Y-\dot\phi(\bX\trans\bbeta)\}^2$ and $\Ubb_{\bbeta}\supm=\{\Hbb_{\bbeta}\supm\}^{-1}$. For simplicity, let $\Hbb\supm_0=\Hbb_{\bbeta_0^{\sf\scriptscriptstyle (m)}}\supm$, $\Jbb_0\supm=\Jbb_{\bbeta_0^{\sf\scriptscriptstyle (m)}}\supm$ and denote by $\u\supm_{0,j}$ the $j\supth$ row of $\Ubb_{\bbeta^{\sf (m)}_0}\supm$. In our following analysis, we assume that the cross--fitting folds $K',K={\Obb(1)}$, $n_m\asymp N/M \equiv n$ for all $m\in[M]$. Here and in the sequel we use $\Obb(1)$ and $\Obb_{\sf P}(1)$ denote of order 1. Next, we introduce assumptions for our theoretical results. For Assumption 4, we only require either 4(a) or 4(b) to hold.

\begin{assumption}[Regular covariance]
(i) There exists absolute constant $C_{\Lambda}>0$ such that for all $m\in[M]$, $C_{\Lambda}^{-1}\leq\Lambda_{\min}(\S_0\supm )\leq\Lambda_{\max}(\S_0\supm )\leq C_{\Lambda}$, $C_{\Lambda}^{-1}\leq\Lambda_{\min}(\Hbb_0\supm )\leq\Lambda_{\max}(\Hbb_0\supm )\leq C_{\Lambda}$ and $C_{\Lambda}^{-1}\leq\Lambda_{\min}(\Jbb_0\supm )\leq\Lambda_{\max}(\Jbb_0\supm )\leq C_{\Lambda}$. (ii) There exist $C_{\Omega}>0$ and $\delta>0$ that for all $m\in[M]$ and $\bbeta\in \Bscr(\bbeta_0\supm,\delta)$, $\ell_1$ norm of each row of $\Ubb_{\bbeta}\supm$ is bounded by $C_{\Omega}$. 
\label{cond:1}
\end{assumption}

\begin{assumption}[Smooth link function]
There exists a constant $C_L>0$ such that for all $\theta,\theta'\in\mathbb{R}$, $|\ddot{\phi}(\theta)-\ddot{\phi}(\theta')|\leq C_L|\theta-\theta'|$.
\label{cond:2}
\end{assumption}

\begin{assumption}[Sub-gaussian residual]
For any $x\in \mathbb{R}^p$, $\epsilon\supm_i$ is conditional sub-gaussian, i.e. there exists $\kappa(x)$ such that $\|\epsilon\supm_i\|_{\psi_2}<\kappa(x)$ given $\bX\supm_i=x$. In addition, there exists some absolute constant $C_{\epsilon}>0$ such that $\kappa(\bX\supm_i)\leq C_{\epsilon}$ and $\ddot{\phi}^{-1}(\bX_i\supmtrans\bbeta\supm_0)\kappa^2(\bX\supm_i)\leq C_{\epsilon}$, almost surely for $m=1,2\ldots,M$.
\label{cond:3}
\end{assumption}

\edef\oldassumption{\the\numexpr\value{assumption}+1}

\setcounter{assumption}{0}
\renewcommand{\theassumption}{\oldassumption(\alph{assumption})}

\begin{assumption}[Sub-gaussian design]
$\bX\supm_i$ is sub-gaussian, i.e. there exists some constant $\kappa>0$ that $\|\bX\supm_i\|_{\psi_2}<\kappa$.
\label{cond:4a}
\end{assumption}
\begin{assumption}[Bounded design]
$\|\bX\supm_i\|_{\infty}$ is bounded by some absolute constant almost surely.
\label{cond:4b} 
\end{assumption}

\begin{remark}
Assumptions \ref{cond:1} (i) and \ref{cond:4a} (or \ref{cond:4b}) are commonly used technical conditions in high-dimensional inference in order to guarantee rate optimality of the regularized regression and debiasing approach \citep{negahban2012unified,javanmard2014confidence}. Assumptions \ref{cond:4a} and \ref{cond:4b} are typically unified by the sub-gaussian design assumption \citep{negahban2012unified}. In our analysis, they are separately studied, since $\|\bX\supm_i\|_{\infty}$ affects the bias rate, which leads to different sparsity assumptions under different design types. Similar conditions as our Assumption \ref{cond:1} (ii) were used in the context of high dimensional precision matrix estimation \citep{cai2011constrained} and debiased inference \citep{chernozhukov2018double,caner2018high,belloni2018high}. Compared with their exact or approximate sparsity assumption imposed on the inverse hessian, this $\ell_1$ boundness assumption is essentially less restrictive. As an important example in our analysis, logistics model satisfies the smoothness conditions for $\phi(\cdot)$ presented by Assumption \ref{cond:2}. As used in \cite{lounici2011oracle} and \cite{huang2010benefit}, Assumption \ref{cond:3} regularizes the tail behavior of the residuals and is satisfied in many common settings like logistic model.

\label{rem:cond}
\end{remark}


\def\half{\frac{1}{2}}
\def\suphalf{^{\half}}

\subsection{Asymptotic properties of the debiased estimator}\label{sec:thm:deb}
We next study the asymptotic properties of the group effect statistics $\zetabreve_j$, $j\in \cH$. We shall 
begin with some important prerequisite results on the convergence properties of $\bbetatilde\supbullet\subnbk$ and 
the debiased estimators $\{\betabreve_j\supm, j\in \cH, m\in [M]\}$ as detailed in Lemmas \ref{lem:1} and \ref{lem:3}.
\begin{lemma}
Under Assumptions \ref{cond:1}-\ref{cond:3}, \ref{cond:4a} or \ref{cond:4b}, and additionally assume that $s=o\{n(\log p)^{-1}\}$, there exist a sequence of the tuning parameters
\[
\lambda\supm_n\asymp \frac{(\log p)\suphalf}{n\suphalf}\quad\mbox{and}\quad\lambda_N\asymp  \frac{\left(M+\log p\right)\suphalf}{n\suphalf M}+\frac{sM^{-\frac{1}{2}}(\log pN)^{a_0}\log p}{n},
\]
with $a_0=1/2$ under Assumption \ref{cond:4a} and $a_0=0$ under Assumption \ref{cond:4b},
such that, for each $k\in [K]$, the integrative estimator satisfies
\[
\|\bbetatilde\supbullet\subnbk-\bbeta_0\supbullet\|_{2,1}=O_{\sf P}(sM\lambda_N), \quad \mbox{and}\quad \|\bbetatilde\supbullet\subnbk-\bbeta_0\supbullet\|_{2}^2=O_{\sf P}(sM^2\lambda^2_N).
\] 
\label{lem:1}
\end{lemma} 
\begin{remark}
Lemma \ref{lem:1} provides the estimation rates of the integrative estimator $\bbetatilde\supbullet\subnbk$. In contrast to the individual--level meta--analysis method, the second term  in the expression of $\lambda_N$ quantifies the additional noise incurred by using summary data under the DataSHIELD constraint. Similar results can be observed through debiasing truncation in distributed learning \citep{lee2017communication,battey2018distributed} or integrative estimation under DataSHIELD \citep{cai2019privacy}. When $s=o\{n^{1/2}(\log pN)^{-a_0}(M+\log p)^{-1}(\log p)^{-1/2}\}$ as assumed in Lemma \ref{lem:3}, the above mentioned error term becomes negligible and our estimator $\bbetatilde\supbullet\subnbk$ achieves the optimal minimax rate in terms of $\ell_2/\ell_1$ and $\ell_2$ errors \citep{lounici2011oracle,huang2010benefit}, which is lower than the local estimation rate as $M$ diverges. Note that, we proposed to send $\bbetatilde\supbullet\subnbk$ back to each data computer and use them to derive the data for debiasing. This in turn contributes to the lower bias rate of our method compared to the one--shot approach.
\label{rem:lem1}
\end{remark}
We next present the theoretical properties of the group debiased estimators.
\begin{lemma}
Under the same assumptions of Lemma \ref{lem:1} and further assume that  
\[
s=o\left\{\frac{n^{\frac{1}{2}}}{(\log pN)^{a_0}(M+\log p)(\log p)^{\frac{1}{2}}}\wedge\frac{n}{M^4(\log p)^4(M+\log p)}\right\},
\]
we have $\betabreve_j\supm-\beta_{0,j}\supm=V_j\supm+\Delta\supm_j$ with $V_j\supm=K^{-1}\sum_{k=1}^K\Pscrhat_{\Isc\supm_k}\u_{0,j}\supmtrans\bX\epsilon$ converging to a normal random variable with mean $0$ and variance $n_m^{-1}(\sigma\supm_{0,j})^2$, where $(\sigma\supm_{0,j})^2 =  {\u_{0,j}\supmtrans\Jbb_0\supm \u_{0,j}\supm}$. In addition, there exists 
$
\tau\asymp {(M+\log p)^{{1}/{2}}}{n^{-{1}/{2}}}
$
such that, simultaneously for all $j\in \cH$, the bias term $\Delta\supm_j$ and the variance estimator $(\sigmahat\supm_j)^2$ satisfy that
\[
|\Delta\supm_j|\leq\sum_{m=1}^M|\Delta\supm_j|=o_{\sf P}\left\{(n\log p)^{-\frac{1}{2}}\right\} 
\quad\mbox{and}\quad\left|(\sigmahat\supm_j)^2-(\sigma\supm_{0,j})^2\right|=o_{\sf P}\left\{(\log p)^{-1}\right\}.
\]
\label{lem:3}
\end{lemma}

\begin{remark}
The sparsity assumption in Lemma \ref{lem:3} is  weaker than the existing debiased estimators for generalized linear models where $s$ is only allowed to diverge in a rate dominated by $N^{\frac{1}{3}}$  \citep{jankova2016confidence,belloni2018high,caner2018high}. This is benefited from the cross--fitting technique, through which we could get rid of the dependence on the convergence rate of $\|\u_{0,j}\supm-\uhat\supm_{j,[k]}\|_1$.
\label{rem:lem3}
\end{remark}
Finally, we establish in Theorem \ref{thm:1} the main result of this section regarding to the asymptotic distribution of the group test statistic $\zetabreve_j$ under the null.  
\begin{theorem}
\label{thm:1}
Under all assumptions in Lemma \ref{lem:3}, simultaneously for all $j\in \cH_0$, we have $\zetabreve_j=S_j+o_{\sf P}(1)$, where $S_j=\sum_{m=1}^Mn_m[V_j\supm/\sigma\supm_{0,j}]^2$. Furthermore, if $M\leq C \log p$ and $\log p = o(N^{1/C'})$ for some constants $C>0$ and $C'>7$, we have
\[
\sup_t|{\sf P}(S_j\leq t)-{\sf P}(\chi^2_M\leq t)|\rightarrow0,~\mbox{as }n,p\rightarrow \infty.
\]
\end{theorem}
The above theorem shows that, the group effect test statistics $\zetabreve_j$ is asymptotically chi-squared distributed under the null and its bias is uniformly negligible for $j\in \cH_0$.

\subsection{False discovery control}\label{sec:thm:fdr}
We establish theoretical guarantees for the error rate control of the multiple testing procedure described in Section \ref{sec:method:mult} in the following two theorems. 

\bet\label{FDR1}
Assume that $q_0= |\mathcal{H}_0|\asymp q$. Then under all assumptions in Lemma \ref{lem:3} with $p\leq cN^r$ and
$M\leq C \log p$ for some constants $c,r,C \in (0,\infty)$, we have $\limsup_{(N,p)\rightarrow \infty}{\text{FDR}(\hat{t})}\leq \alpha$, and for any $\epsilon>0$, $\lim_{(N,p)\rightarrow \infty}\pr\{\text{FDP}(\hat{t})\leq \alpha+\epsilon\}=1$.
\eet
\begin{remark}
\label{correlation}
Assumption \ref{cond:1} (i) ensures that most of the group estimates $\{\zetabreve_j, j\in \cH_0\}$ are not highly correlated with each other. Thus the the variance of $\Rhat_0(t)$ can be appropriately controlled, which in turn guarantees the control of false discovery proportion. 
\end{remark}

As described in Section \ref{sec:method:mult}, if $\hat{t}$ in equation \eqref{t_hat} is not attained in the range $[0, \; (2\log q-2\log\log q)^{1/2}]$, then it is thresholded  at $(2\log q)^{1/2}$. The following theorem states a weak condition to ensure the existence of $\hat{t}$ in such range. As a result, the false discovery proportion and false discovery rate will converge to the pre-specified level $\alpha$ asymptotically.
\bet\label{FDR2}
Let $\mathcal{S}_\rho= \left\{j\in \cH: \sum_{m=1}^Mn_m [\beta_{0,j}\supm]^2 \geq (\log q)^{1+\rho}\right\}$. Suppose for some $\rho>0$ and some $\delta>0$, $|\mathcal{S}_\rho| \geq \{{1}/({\pi}^{1/2}\alpha)+\delta\}({\log q})^{1/2}$.
Then under the same conditions as in Theorem \ref{FDR1}, we have, as $(N,p)\rightarrow \infty$,
\[
\frac{\text{FDR}(\hat{t})}{\alpha q_0/q} \rightarrow 1, \quad \frac{\text{FDP}(\hat{t})}{\alpha q_0/q} \rightarrow 1 \mbox{ in probability.}
\]
\eet
In the above theorem, the condition on $\mathcal{S}_\rho$ only requires very few covariates having the signal sum of squares across the studies $\sum_{m=1}^M [\beta_{0,j}\supm]^2$ exceeding the rate $(\log q)^{1+\rho}/n_m$ for some $\rho>0$, and is thus a very mild assumption.

\subsection{Comparison with alternative approaches}\label{sec:thm:com}
To study the advantage of our testing approach and the impact of the DataSHIELD constraint, we next compare the proposed method to a one--shot approach and the individual--level meta--analysis approach, as described in Algorithms \ref{alg:one-shot} and \ref{alg:ilma},  through a theoretical perspective. 
The one--shot approach in Algorithm \ref{alg:one-shot} is inspired by existing literature in distributed learning  \citep[e.g.]{lee2017communication,battey2018distributed} {and is a natural extension of existing methods to the problem of multiple testing under the DataSHIELD constraint. The debiasing step of the one--shot approach is performed locally as in the existing literatures.} 

\begin{algorithm}[htbp]
\caption{\label{alg:one-shot} One--shot approach.}
\begin{enumerate}[Step 1.]
\item[Step 1.] \label{alg2:step1} At each data computer, obtain the cross--fitted debiased estimator by solving a dantzig selector problem locally, where $\bbeta\supm$ is estimated by local LASSO.

\item[Step 2.] \label{alg2:step2} Send the debiased estimators to the analysis computer and obtain the group statistics. 

\item[Step 3.] \label{alg2:step3} Perform multiple testing procedure as described in Section \ref{sec:method:mult}.
\end{enumerate}
\end{algorithm}

\begin{algorithm}[htbp]
\caption{\label{alg:ilma} Individual--level meta--analysis.}
\begin{enumerate}[Step 1.]
\item[Step 1.] \label{alg3:step1} Integrate all individual--level data at the analysis computer.

\item[Step 2.] \label{alg3:step2} Construct the cross--fitted debiased estimator by (\ref{equ:deb}) using individual--level integrative estimator analog to (\ref{equ:adele}), and then obtain the overall effect statistics.

\item[Step 3.] \label{alg3:step3} Perform multiple testing procedure in Section \ref{sec:method:mult}.

\end{enumerate}
\end{algorithm}

Parallel to the proofs of Lemma \ref{lem:3} and Theorems \ref{FDR1} and \ref{FDR2},  in order to achieve the same error rate control results, the sparsity assumptions for the one--shot and the individual--level meta--analysis approaches are respectively $s=o\left\{{n^{{1}/{2}}}{(\log pN)^{-a_0}M^{-1}(\log p)^{-{3}/{2}}}\wedge{n}{M^{-4}(\log p)^{-5}}\right\}$ and $s=o\left\{{n^{{1}/{2}}}{(\log pN)^{-a_0}(M+\log p)^{-1}(\log p)^{-{1}/{2}}}\wedge{n}{M^{-4}(\log p)^{-4}(M+\log p)^{-1}}\right\}$, while the other assumptions remain to be the same. In addition, if $p\leq cN^r$ and $M\leq C\log p$ as assumed in Theorems \ref{FDR1} and \ref{FDR2}, the sparsity conditions for these two approaches reduce to
\[
s=o\left\{n^{\frac{1}{2}}(\log pN)^{-a_0}M^{-1}(\log p)^{-\frac{3}{2}}\right\}\quad\mbox{and}\quad s=o\left\{n^{\frac{1}{2}}(\log pN)^{-a_0}(\log p)^{-\frac{1}{2}}\right\},
\]
respectively. Our method requires the same sparsity assumption as the individual--level meta--analysis approach. This is because the additional error rate of the proposed method is negligible under our sparsity assumption, as discussed in Remark \ref{rem:lem1}. When $M$ diverges with $n$ and $p$, the individual--level meta--analysis approach and the proposed method require strictly weaker assumption than the one--shot approach, as shown in Table \ref{tab:rate}.

\begin{table}
\small
\centering
\small \caption{\label{tab:rate} Sparsity assumptions under the bounded and sub-Gaussian design with $p\leq cN^r$ and $M\leq C\log p$, where $\gamma_1=(n/\log p)^{1/2}(\log p)^{-1}$, 
$\gamma_2=(n/\log p)^{1/2} (M\log p)^{-1}$.}
\def\arraystretch{1.3}
\begin{tabular}{|c|c|c|c|}
\hline
    & Our method & individual--level meta--analysis & One--shot  \\ \hline
Bounded & $s=o(\gamma_1)$  &  $s=o(\gamma_1)$   & $s=o(\gamma_2)$ \\ \hline
Sub-Gaussian  & $s=o\{\gamma_1(\log pN)^{-{1}/{2}}\}$  & $s=o\{\gamma_1(\log pN)^{-{1}/{2}}\}$  & $s=o\{\gamma_2(\log pN)^{-{1}/{2}}\}$  \\ \hline
\end{tabular}
\end{table}

\begin{remark} Our approach involves transferring data twice from the data computers to the analysis computer and once from the analysis computer to the data computers, which requires more communication efforts compared to the one--shot approach. The additional communication gains lower bias rate than the one--shot approach while only requiring the same sparsity assumption as the individual--level meta--analysis method, as shown in Table \ref{tab:rate}. Under its sparsity condition, each method is able to draw inference that is asymptotically valid and has the same power as the ideal case when one uses the true parameters in construction of the group test statistics. This further implies that, to construct a powerful and valid multiple testing procedure, there is no necessity to adopt further sequential communications between the data computers and the analysis computer like the distributed methods such as \cite{li2016supporting} and \cite{wang2017efficient}.
\label{rem:comp2}
\end{remark}



\def\expit{\mbox{expit}}

\section{Simulation Study}\label{sec:simu}


We evaluate the empirical performance of the proposed testing procedure and compare it with the one--shot and the individual--level meta--analysis methods. Throughout, we let $M=5$, $n_m=500$, and vary 
$p$ from $500$ to $1000$. For each setting, we perform $200$ replications and set the number of sample splitting folds $K=2$, $K'=5$ and false discovery level $\alpha=0.1$. The tuning strategies described in Section \ref{sec:method:tune} are employed with $H=10$. 

The covariates $\X$ of each study is generated from either (i) Gaussian auto--regressive model of order $1$ and correlation coefficient $0.5$; or 
(ii) Hidden markov model with binary hidden variables and binary observed variables with the transition probability probability and the emission probability both set as $0.2$. We choose $\{\bbeta\supm_0\}$ to be heterogeneous in magnitude across studies but share the same support  with
\[
\bbeta\supm_0=\mu \left\{(\nu_1\supm+1)\psi_1, (\nu_2\supm+1)\psi_2,\ldots,(\nu_s\supm+1)\psi_{s},\mathbf{0}_{p-s}\right\}\trans
\]
where the sparsity level $s$ is set to be $10$ or $50$, $\{\psi_1, ..., \psi_s\}$ are independently drawn from $\{-1,1\}$ with equal probability and are shared across studies while the local signal strength $\nu_j\supm$'s vary across studies and are drawn independently from N$\{0, (\mu/2)^2\}$. To ensure the procedures have reasonable power magnitudes for comparison, we set the overall signal strength $\mu$ to be in the range of $[0.21, 0.42]$ for $s=10$, mimicking a sparse and strong signal setting; and $[0.14, 0.35]$ for $s=50$, mimicking a dense and weak signal setting. 
We then generate  binary responses $Y\supm$ from $\mbox{logit}P(Y\supm = 1 \mid \bX\supm) = \bbeta\supmtrans_0\bX\supm$.

\begin{figure}[htbp]
\centering
\includegraphics[width=1\textwidth]{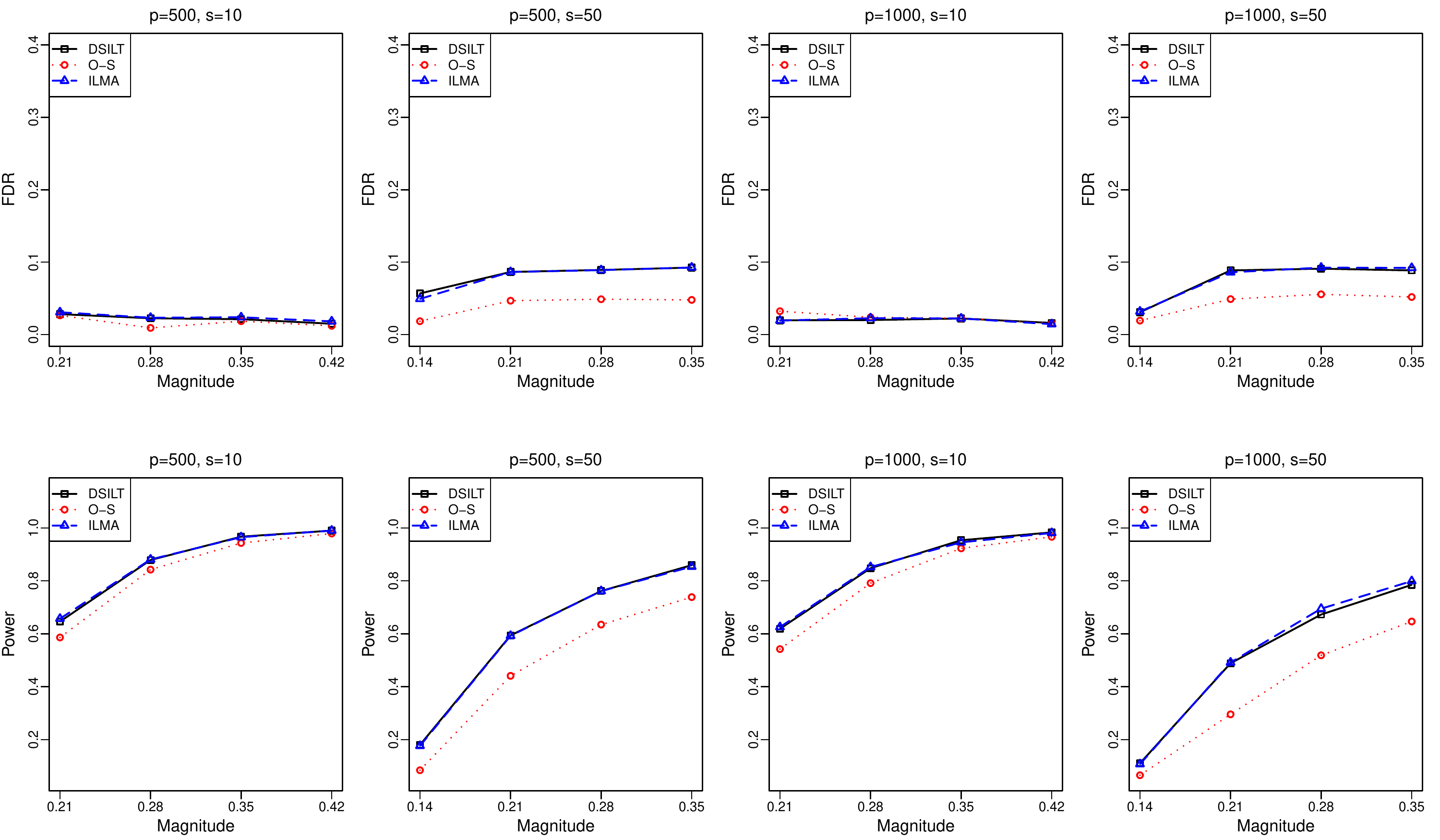}
\caption{The empirical false discovery rate, denote by FDR, and empirical power of our proposed method, the one--shot approach and the individual--level meta--analysis method, denoted respectively by DSILT, O--S and ILMA, under the Gaussian design, with $\alpha=0.1$. The horizontal axis represents the overall signal magnitude $\mu$.} 
\label{fig:set1}
\end{figure}

In Figure \ref{fig:set1}, we report the empirical false discovery rate and power of the three methods with varying $p$, $s$, and $\mu$ 
under the Gaussian design. Results for hidden markov model design has almost the same pattern and are included in the Supplementary Material. Across all settings,  our method achieves almost the same performance as the ideal approach in both error rate control and power. All the methods successfully control the desired false discovery rate at $\alpha=0.1$. When $s=10$ or the signal strength $\mu$ is weak, all the methods have conservative error rates compared to the nominal level. While for $s=50$ with relatively strong signal, our method and the ideal approach become close to the exact error rate control empirically. This is consistent with Theorem \ref{FDR2} that if the number of relatively strong signals is large enough, our method tends to achieve exact false discovery rate  control. 
The difference in empirical power between our method and the individual--level meta--analysis approach is less than $1\%$ in all cases. This indicates that the proposed testing procedure can accommodates the DataSHIELD constraint at almost no cost in power compared to ideal method. This is consistent with our theoretical result in Section \ref{sec:thm:com} that the two methods require the same sparsity assumption for simultaneous inference.

Furthermore, the proposed method and individual--level meta--analysis method dominate the one--shot strategy in terms of statistical power. Under every single scenario, the power of the former two methods is around $15\%$ higher than that of the one--shot approach in the dense case, i.e., $s=50$, and $6\%$ higher in the sparse case, i.e, $s=10$. By developing testing procedures using integrative analysis rather than local estimations, both our method and individual--level meta--analysis method utilize the group sparsity structure of the model parameters $\bbeta\supbullet$ more adequately than the one--shot approach, which leads to the superior power performance of these two methods.  
The power advantage is more pronounced as the sparsity level $s$ grows from $10$ to $50$. This is due to the fact that, to achieve the same result, the one--shot approach requires stronger sparsity assumption than the other two methods, and is thus much easier to be impacted by the growth of $s$. In comparison, the performance of our method and individual--level meta--analysis method is less sensitive to the sparsity growth because the integrative estimator employed in these two methods is more stable than the local estimator under the dense scenario.

\def\submain{_{\scriptscriptstyle \sf main}}
\def\subint{_{\scriptscriptstyle \sf int}}
\def\MVPf{\mbox{MVP}_f}
\def\PHBf{\mbox{PHB}_f}
\def\UKBf{\mbox{UKB}_f}
\def\MVPm{\mbox{MVP}_m}
\def\PHBm{\mbox{PHB}_m}
\def\UKBm{\mbox{UKB}_m}

\section{Real Example}\label{sec:real}

Statins are the most widely prescribed drug for lowering the low--density lipoprotein and the risk of cardiovascular disease with over a quarter of 45 years or older adults receiving this drug in the United States. Statins lower low--density lipoprotein by inhibiting 3-hydroxy-3-methylglutaryl-coenzyme A reductase \citep{nissen2005statin}. The treatment effect of statins can be also causally inferred based on the effect of the 3-hydroxy-3-methylglutaryl-coenzyme A reductase variant \emph{rs17238484} -- patients carrying the \emph{rs17238484}-G allele have profiles similar to individuals receiving statin, with lower low--density lipoprotein and lower risk of cardiovascular disease \citep{swerdlow2015hmg}. 
While the benefit of statin has been consistently observed, it is not without risk. There has been increasing evidence that statins increase the risk of type II diabetes \citep{rajpathak2009statin,carter2013risk}. \cite{swerdlow2015hmg} demonstrated via both meta analysis of clinical trials and genetic analysis of  the  \emph{rs17238484} variant that statins are associated with a slight increase of type II diabetes risk. However, the adverse effect of statin on type II diabetes risk appears to differ substantially depending on the number of type II diabetes risk factors patients have prior to receiving statin, with adverse risk higher among patients with more risk factors \citep{waters2013cardiovascular}. 

To investigate potential genetic determinants of the statin treatment effect heterogeneity, we studied interactive effects of the \emph{rs17238484} variant and 256 SNPs associated with type II diabetes, low--density lipoprotein, high--density lipoprotein cholesterol, and the coronary artery disease gene which plays a central role in obesity and insulin sensitivity \citep{kozak2009ucp1,rodrigues2013genetic}. A significant interaction between SNP $j$ and the statin variant \emph{rs17238484} would indicate that SNP $j$ modifies the effect of statin. Since the low--density lipoprotein, coronary artery disease and type II diabetes risk profiles differ greatly between different racial groups and between male and female, we focus on the analysis on the black sub-population and fit separate models for female and male subgroups. 

To efficiently identify genetic risk factors that significantly interact with \emph{rs17238484}, we performed an integrative analysis of data from 3 different studies, including the Million Vetern Project from Veteran Health Administration, Partners Healthcare Biobank and UK Biobank. Within each study, we have both a male subgroup indexed by subscript $m$, and a female subgroup indexed by subscript $f$, leading to $M=6$ datasets denoted by $\MVPf, \MVPm, \PHBf, \PHBm, \UKBf$ and $\UKBm$. Since type II diabetes prevalence within the datasets varies greatly from 0.05\% to 0.15\%, we performed a case control sampling with 1:1 matching so each dataset has equal numbers of type II diabetes cases and controls. Since Million Vetern Project has a substantially larger number of male type II diabetes cases than all other studies, we down sampled its cases to match the number of female cases in Million Vetern Project so that the signals are not dominated by the male population.  This leads to sample sizes of 216, 392, 606, 822, 3120 and 3120 at PHB$_m$, PHB$_f$, UKB$_m$, UKB$_f$, MVP$_m$ and MVP$_f$, respectively. The covariate vector $\X = (\X\submain\trans, \X\subint\trans)\trans$ is of dimension $p=516$, where $\X\submain$ consists of the main effects of \emph{rs17238484}, age and the aforementioned 256 SNPs, and $\X\subint$ consists of the interactions between \emph{rs17238484} and age as well as each of the 256 SNPs. All SNPs are encoded such that higher value is associated with higher risk of type II diabetes. We implemented the proposed testing method along with the one--shot approach as a benchmark to perform multiple testing of $q=256$ coefficients corresponding to the interaction terms in $\X\subint$ at nominal level of $\alpha = 0.1$ with the model chosen as logistics regression and the sample splitting folds $K=2$ and $K'=5$. 

\begin{figure}[htbp]
\centering
\includegraphics[width = 1.88in]{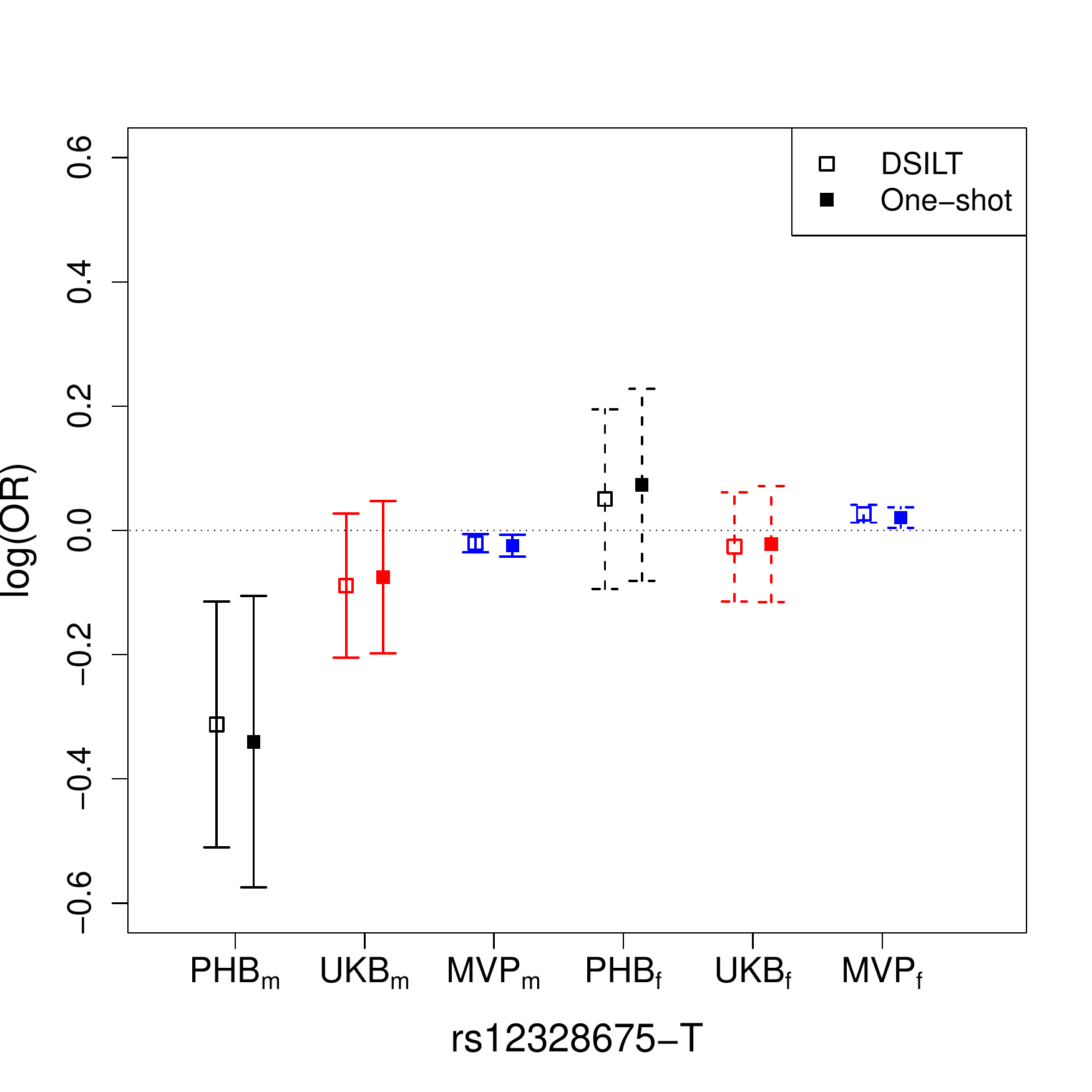}
\includegraphics[width = 1.88in]{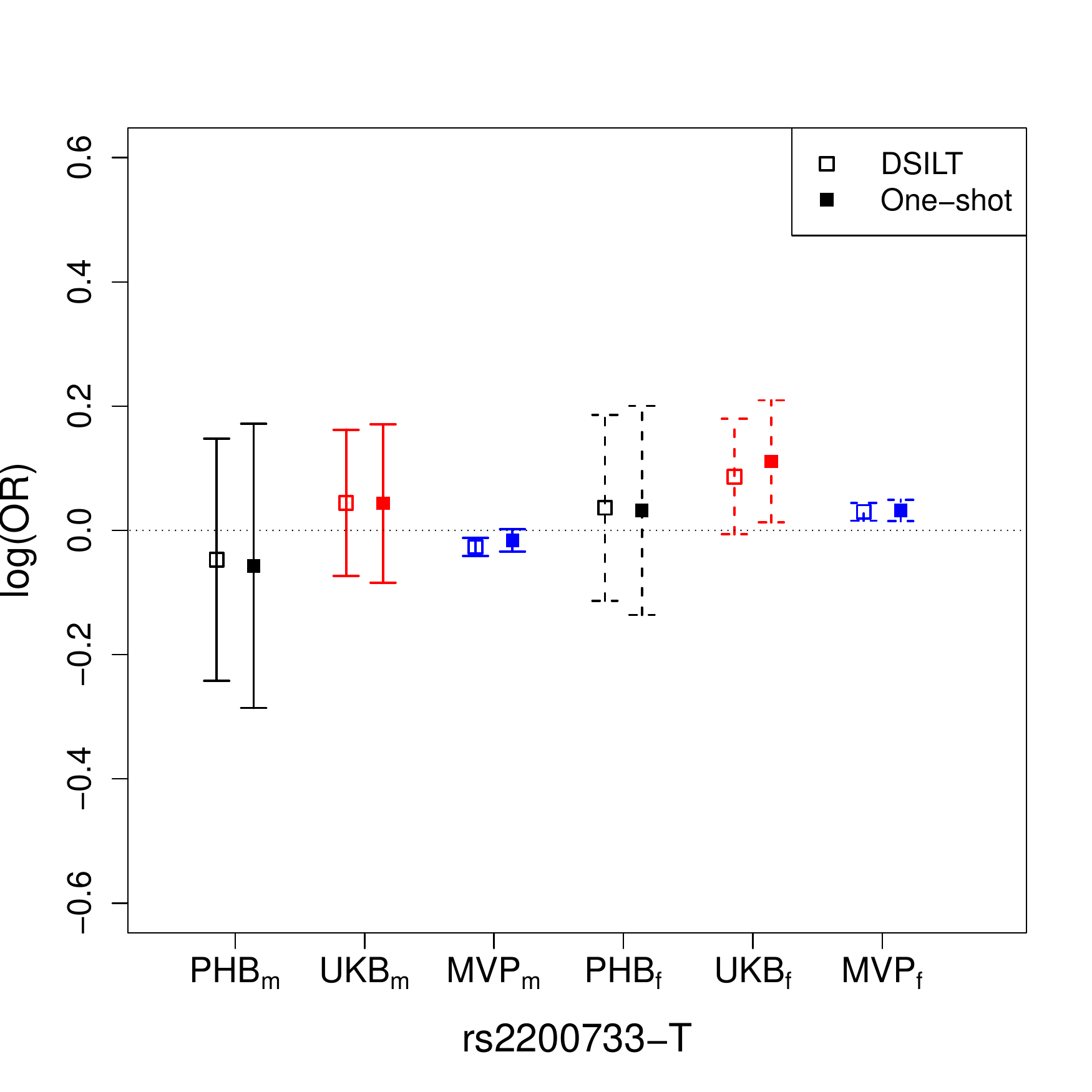}
\includegraphics[width = 1.88in]{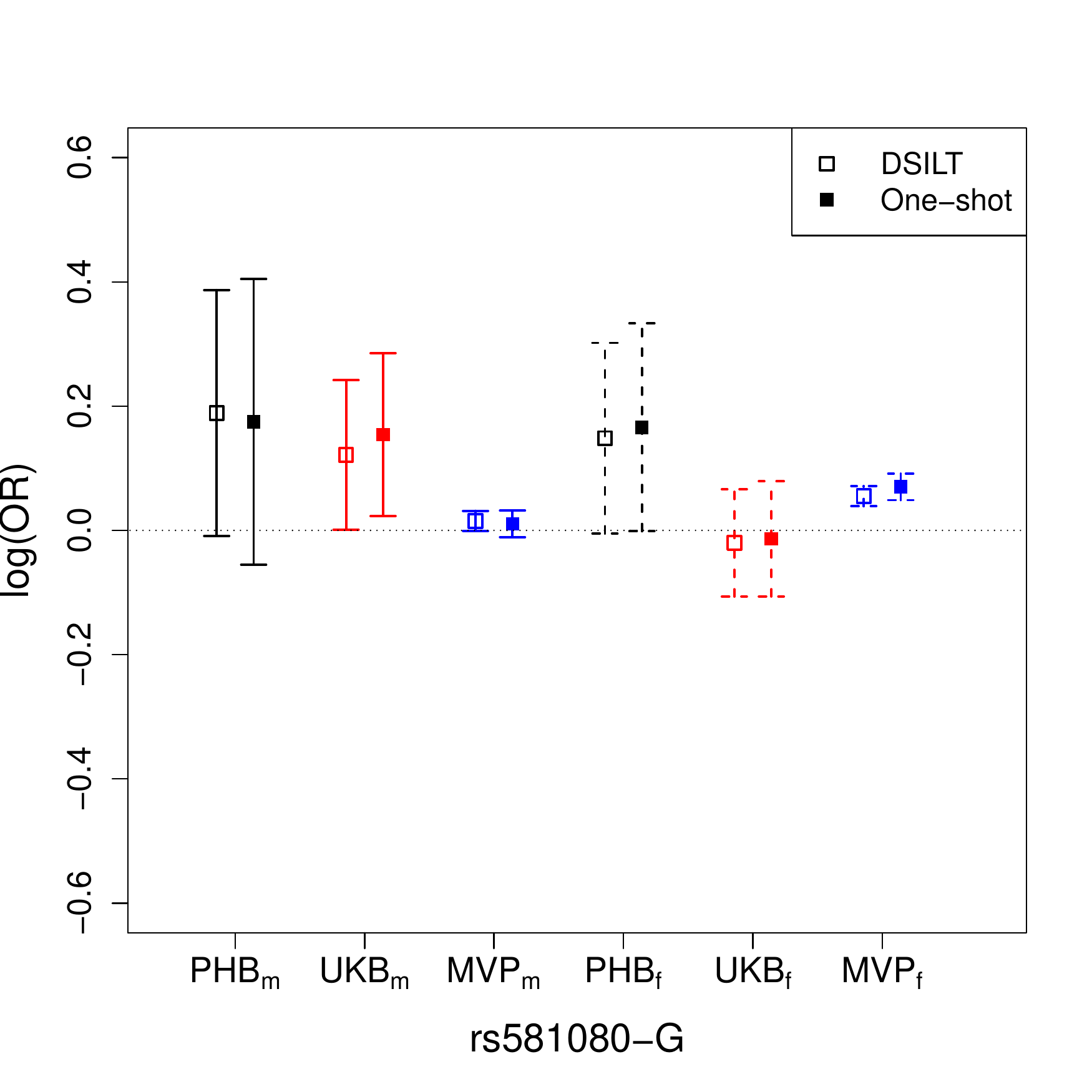}
\includegraphics[width = 1.88in]{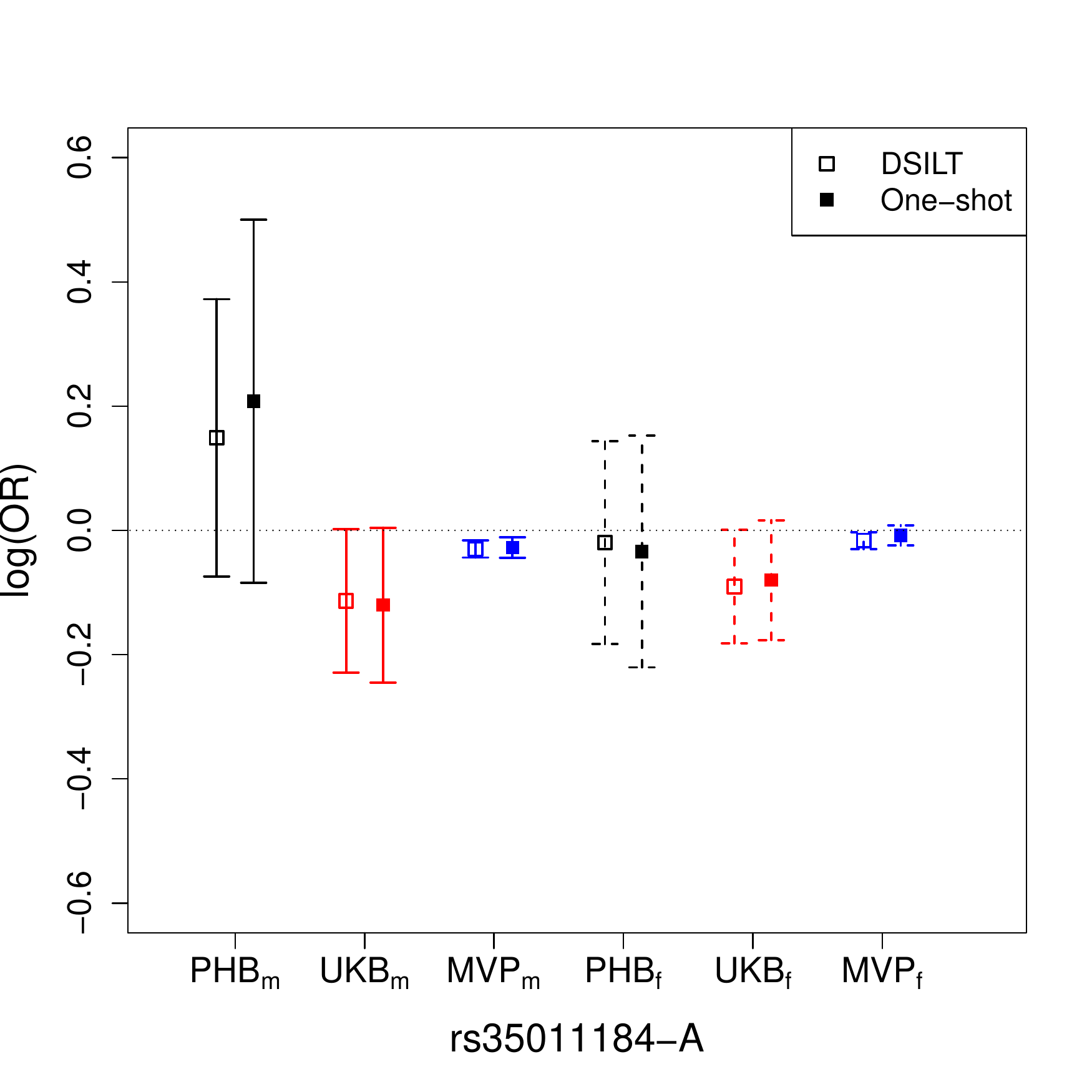}
\includegraphics[width = 1.88in]{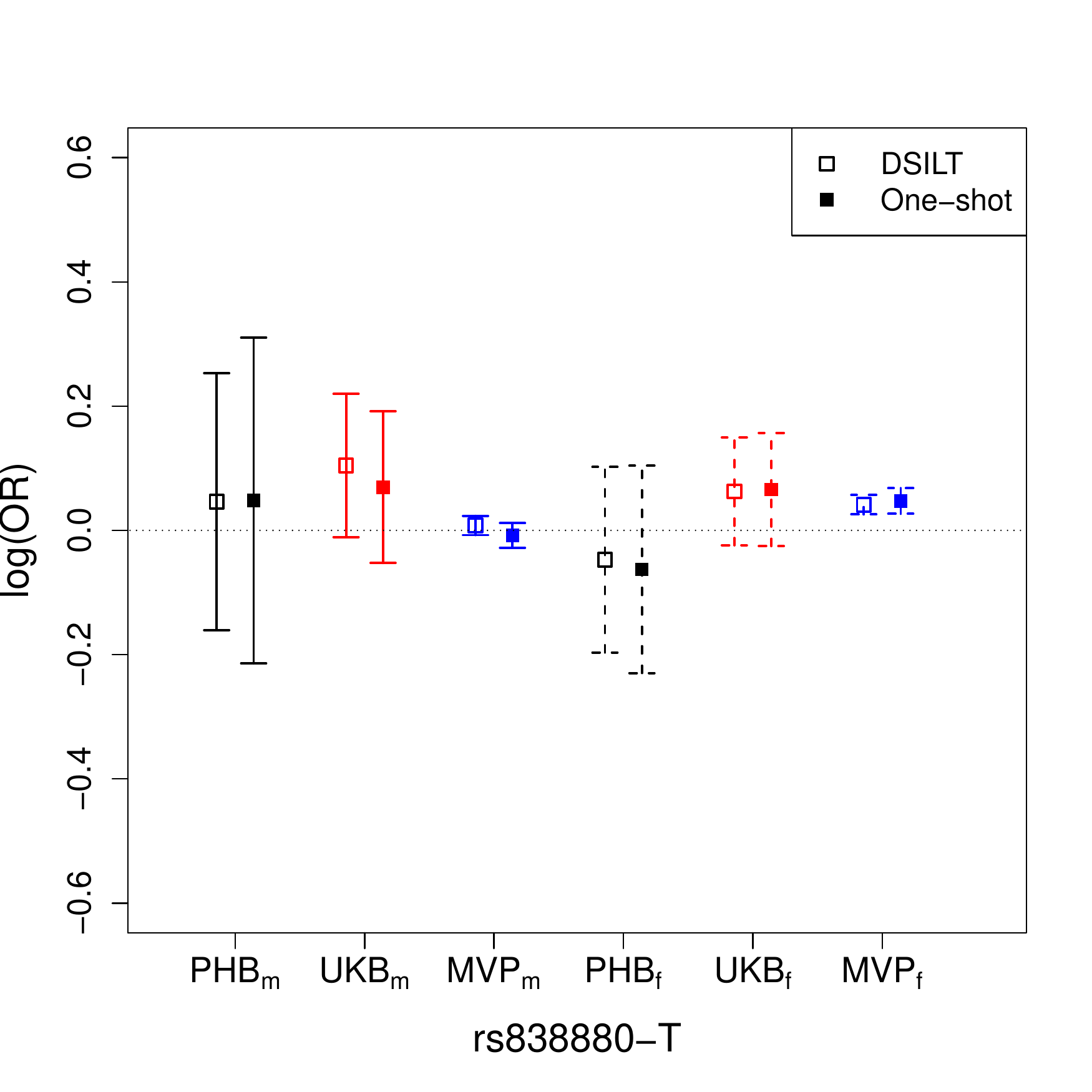}
\caption{\label{beta:results} Debiased estimates of the log odds ratios and their 90\% confidence intervals in each local site for the interaction effects between \emph{rs17238484}-G and the 5 SNPs detected by our method, obtained respectively based on the one--shot and the proposed approach, denoted by DSILT.}
\end{figure}

As shown in Table \ref{tab:results}, our method identifies 5 SNPs significantly interacting with the statin SNP while the one--shot approach detects only 3 SNPs, all of which belong to the set of SNPs identified by our method.  The presence of non-zero interactive effects demonstrates that the adverse effect of statin SNP \emph{rs17238484}-G on the risk of type II diabetes can differ significantly among patients with different levels of genetic predisposition to type II diabetes. In Figure \ref{beta:results}, we also present 90\% confidence intervals obtained within each dataset for the interactive effects between \emph{rs17238484}-G and each of these 5 detected SNPs. The SNP {\em rs581080}-G in the TTC39B gene has the strongest interactive effect with the statin SNP and has all interactive effects estimated as positive for most studies, suggesting that the adverse effect of statin is generally higher for patients with this mutation compared to those without. Interestingly, there has been a previous report suggesting that a SNP in the TTC39B  gene is associated with statin induced response to low--density lipoprotein particle number \citep{chu2015differential}, suggesting that the effect of statin can be modulated by the {\em rs581080}-G SNP. 

Results shown in Figure \ref{beta:results} also suggest some gender differences in the interactive effects. For example, the adverse effect of statin is lower for female patients carrying the {\em rs12328675-T} allele compare to female patients without the allele. On the other hand, the effect of statin appear to be higher for male patients with the {\em rs12328675-T} allele compared to those without genetic variants associated with a various of phenotypes related to type II diabetes. The variation in the effect sizes across different data sources illustrates that it is necessary to properly account for heterogeneity of $\bbeta$ in the modeling procedure. Comparing the lengths of confidence intervals obtained based on the one--shot approach to those from the proposed method, we find that the data shielding integrative approach generally yields shorter confidence intervals, which translates to higher power in signal detection.  It is important to note that since Million Vetern Project has much larger sample sizes, the width of the confidence intervals from Million Vetern Project are much smaller than those of UK Biobank and Partners Healthcare Biobank. However, the effect sizes obtained from Million Vetern Project also tend to be much smaller in magnitude and consequently, using Million Vetern Project alone would only detect 2 of all these 5 SNPs by multiple testing with level $0.1$. This demonstrates the utility of the integrative testing involving $M=6$ data sources.

\begin{table}[htbp]
\small
\caption{SNPs identified by our method to interact with the statin genetic variants \emph{rs17238484}-G on the risk for type II diabetes. 
The second column presents the name of the gene where the SNP locates. The third column presents the minor allele frequency, denoted by MAF, of each SNP averaged over the three sites. The last three columns respectively present the $p$--values obtained using one--shot approach with all the $M=6$ studies, one--shot with solely the datasets ${\rm MVP}_f$ and ${\rm MVP}_m$ and the proposed method with all the $M=6$ studies. The $p$--values shown in black fonts represent the SNPs selected by each method.}
\label{tab:results} 
\centering
\begin{tabular}{cccccc}
\hline
SNP          & Gene & MAF & One--shot & MVP--only & Our method \\ \hline 
\emph{rs12328675}-T   &  	COBLL1   &   $0.13$    & ${\bf 1.1\times 10^{-3}}$   &  $2.3\times10^{-3}$    & ${\bf 6.0\times 10^{-4}}$ \\ \hline 
\emph{rs2200733}-T  & 	LOC729065      &   $0.18$ & $3.7\times 10^{-2}$  &  $5.7\times10^{-3}$  & ${\bf 6.2\times 10^{-4}}$ \\ \hline 
\emph{rs581080}-G &  TTC39B            &   $0.22$   & ${\bf 3.6\times 10^{-6}}$     &  ${\bf 1.1\times10^{-6}}$    & ${\bf 2.6\times 10^{-6}}$ \\ \hline 
\emph{rs35011184}-A  & TCF7L2           &   $0.22$      & $1.9\times 10^{-2}$   &  $5.2\times10^{-2}$  & ${\bf 8.6\times 10^{-4}}$ \\ \hline 
\emph{rs838880}-T  & SCARB1    &  $0.36$  & ${\bf 6.7\times 10^{-4}}$    &    ${\bf 6.0\times 10^{-5}}$    & ${\bf 6.2\times 10^{-4}}$ \\ \hline 
\end{tabular}
\end{table}

\section{Discussion}\label{sec:dis}
In this paper, we propose a data shielding integrative large-scale testing method for simultaneous inference of high dimensional covariate effects in the presence of between study heterogeneity under the DataSHIELD framework. The proposed method is able to properly control the false discovery rate and false discovery proportion in theory asymptotically, and is shown to have similar performance as the ideal individual--level meta--analysis method and outperform the one--shot approach in terms of the required assumptions and the statistical power for multiple testing.
We demonstrate  that the sparsity assumptions of the proposed method are equivalent to those for the ideal method but strictly weaker than those for the one--shot approach. As a price to pay, our method requires one more round of data transferring between the analysis computer and the data computers than the one--shot approach. Meanwhile, the sparsity condition equivalence between the proposed method and individual--level meta--analysis method implies that there is no need to include in our method further rounds of communications or adopt iterative procedures like \cite{li2016supporting} and \cite{wang2017efficient}, which saves lots of human efforts in practice.

The proposed approach also adds technical contributions to existing literatures in several aspects. First, our debiasing formulation helps to get rid of the group structure assumption on the covariates $\bX\supm$ at different distributed sites. Such assumption is not satisfied in our real data setting, but is unavoidable if one uses the node-wise group LASSO \citep{mitra2016benefit} or group structured inverse regression \citep{xia2018joint} for debiasing. Second, compared with the existing work on joint testing of high dimensional linear models \citep{xia2018joint}, our method considers model heterogeneity and allows the number of studies $M$ to diverge under the data sharing constraint, resulting substantial technical difficulties in characterizing the asymptotic distribution of our proposed test statistics $\breve{\zeta}_j$ and their correlation structures for the simultaneous inference. 

We next discuss the limitation and possible extension of the current work. First, the proposed procedure requires transferring of Hessian matrix with $O(p^2)$ complexity from each data computer to the analysis computer. To the best of our knowledge, there is no natural way to reduce such order of complexity for the group debiasing step, i.e., Step \ref{alg1:step2}, as introduced in Section \ref{sec:method:de}. Nevertheless, it is worthwhile to remark that, for the integrative estimation step, i.e., Step \ref{alg1:step1}, the communication complexity can be reduced to $O(p)$ only, by first transferring the locally debiased LASSO estimators from each data computer to the analysis computer and then integrating the debiased estimators with a group structured truncation procedure \citep[e.g.]{lee2017communication,battey2018distributed} to obtain an integrative estimator with the same error rate as $\bbetatilde\supbullet_{[\text{-}k]}$. However, such procedure requires more efforts in deriving the data at each data computer, which is not easily attainable in some situations such as our real example. 
{Second, we assume $q=|\cH| \asymp p$ in the current paper as we have $q=p/2$ in the real example of Section \ref{sec:real}. We can further extend our results to the cases when $q$ grows slower than $p$. In such scenarios, the error rates control results in Theorems \ref{FDR1} and \ref{FDR2} still hold, and in the mean while the model sparsity assumptions and the conditions on $p$ and $N$ can be further relaxed because we have fewer number of hypotheses to test in total and as a result the error rate tolerance for an individual test $H_{0,j}$ can be weakened.}
Third, we require $M=O(\log p)$ for the limiting null distribution of the test statistics $\breve{\zeta}_j$ and the subsequent simultaneous error rates control.
Such assumption is naturally satisfied in many situations as in our real example. However, when the collaboration is of a larger scale, say $M\gg \log p$ or $M>n_m$, developing an adaptive and powerful overall effect testing procedure, particularly under DataShield constraints, warrants future research.




\section*{Supplemental Materials}
Supplementary materials available at Biometrika online include the proofs, technical lemmas and additional simulation results. 



\bibliographystyle{biometrika}
\bibliography{library}

\begin{thebibliography}{50}
\expandafter\ifx\csname natexlab\endcsname\relax\def\natexlab#1{#1}\fi

\bibitem[{Allen et~al.(2002)Allen, Bourhis, Burrell \&
  Mabry}]{allen2002comparing}
\textsc{Allen, M.}, \textsc{Bourhis, J.}, \textsc{Burrell, N.} \&
  \textsc{Mabry, E.} (2002).
\newblock Comparing student satisfaction with distance education to traditional
  classrooms in higher education: A meta-analysis.
\newblock \textit{The American Journal of Distance Education} \textbf{16},
  83--97.

\bibitem[{Battey et~al.(2018)Battey, Fan, Liu, Lu, Zhu
  et~al.}]{battey2018distributed}
\textsc{Battey, H.}, \textsc{Fan, J.}, \textsc{Liu, H.}, \textsc{Lu, J.},
  \textsc{Zhu, Z.} et~al. (2018).
\newblock Distributed testing and estimation under sparse high dimensional
  models.
\newblock \textit{The Annals of Statistics} \textbf{46}, 1352--1382.

\bibitem[{Belloni et~al.(2018)Belloni, Chernozhukov, Chetverikov, Hansen \&
  Kato}]{belloni2018high}
\textsc{Belloni, A.}, \textsc{Chernozhukov, V.}, \textsc{Chetverikov, D.},
  \textsc{Hansen, C.} \& \textsc{Kato, K.} (2018).
\newblock High-dimensional econometrics and regularized gmm.
\newblock \textit{arXiv preprint arXiv:1806.01888} .

\bibitem[{Bradfield et~al.(2012)Bradfield, Taal, Timpson, Scherag, Lecoeur,
  Warrington, Hypponen, Holst, Valcarcel, Thiering
  et~al.}]{bradfield2012genome}
\textsc{Bradfield, J.~P.}, \textsc{Taal, H.~R.}, \textsc{Timpson, N.~J.},
  \textsc{Scherag, A.}, \textsc{Lecoeur, C.}, \textsc{Warrington, N.~M.},
  \textsc{Hypponen, E.}, \textsc{Holst, C.}, \textsc{Valcarcel, B.},
  \textsc{Thiering, E.} et~al. (2012).
\newblock A genome-wide association meta-analysis identifies new childhood
  obesity loci.
\newblock \textit{Nature genetics} \textbf{44}, 526.

\bibitem[{Cai et~al.(2019{\natexlab{a}})Cai, Li, Ma \&
  Xia}]{cai2019differential}
\textsc{Cai, T.}, \textsc{Li, H.}, \textsc{Ma, J.} \& \textsc{Xia, Y.}
  (2019{\natexlab{a}}).
\newblock Differential markov random field analysis with an application to
  detecting differential microbial community networks.
\newblock \textit{Biometrika} \textbf{106}, 401--416.

\bibitem[{Cai et~al.(2019{\natexlab{b}})Cai, Liu \& Xia}]{cai2019privacy}
\textsc{Cai, T.}, \textsc{Liu, M.} \& \textsc{Xia, Y.} (2019{\natexlab{b}}).
\newblock Individual data protected integrative regression analysis of
  high-dimensional heterogeneous data.
\newblock \textit{arXiv preprint arXiv:1902.06115} .

\bibitem[{Cai et~al.(2011)Cai, Liu \& Luo}]{cai2011constrained}
\textsc{Cai, T.}, \textsc{Liu, W.} \& \textsc{Luo, X.} (2011).
\newblock A constrained l1 minimization approach to sparse precision matrix
  estimation.
\newblock \textit{Journal of the American Statistical Association}
  \textbf{106}, 594--607.

\bibitem[{Caner \& Kock(2018)}]{caner2018high}
\textsc{Caner, M.} \& \textsc{Kock, A.~B.} (2018).
\newblock High dimensional linear gmm.
\newblock \textit{arXiv preprint arXiv:1811.08779} .

\bibitem[{Card et~al.(2010)Card, Kluve \& Weber}]{card2010active}
\textsc{Card, D.}, \textsc{Kluve, J.} \& \textsc{Weber, A.} (2010).
\newblock Active labour market policy evaluations: A meta-analysis.
\newblock \textit{The economic journal} \textbf{120}, F452--F477.

\bibitem[{Carter et~al.(2013)Carter, Gomes, Camacho, Juurlink, Shah \&
  Mamdani}]{carter2013risk}
\textsc{Carter, A.~A.}, \textsc{Gomes, T.}, \textsc{Camacho, X.},
  \textsc{Juurlink, D.~N.}, \textsc{Shah, B.~R.} \& \textsc{Mamdani, M.~M.}
  (2013).
\newblock Risk of incident diabetes among patients treated with statins:
  population based study.
\newblock \textit{Bmj} \textbf{346}, f2610.

\bibitem[{Chernozhukov et~al.(2016)Chernozhukov, Chetverikov, Demirer, Duflo,
  Hansen \& Newey}]{chernozhukov2016double}
\textsc{Chernozhukov, V.}, \textsc{Chetverikov, D.}, \textsc{Demirer, M.},
  \textsc{Duflo, E.}, \textsc{Hansen, C.} \& \textsc{Newey, W.~K.} (2016).
\newblock Double machine learning for treatment and causal parameters.
\newblock Tech. rep., cemmap working paper.

\bibitem[{Chernozhukov et~al.(2018)Chernozhukov, Newey \&
  Robins}]{chernozhukov2018double}
\textsc{Chernozhukov, V.}, \textsc{Newey, W.} \& \textsc{Robins, J.} (2018).
\newblock Double/de-biased machine learning using regularized riesz
  representers.
\newblock \textit{arXiv preprint arXiv:1802.08667} .

\bibitem[{Chu et~al.(2015)Chu, Giulianini, Barratt, Ding, Nyberg, Mora, Ridker
  \& Chasman}]{chu2015differential}
\textsc{Chu, A.~Y.}, \textsc{Giulianini, F.}, \textsc{Barratt, B.~J.},
  \textsc{Ding, B.}, \textsc{Nyberg, F.}, \textsc{Mora, S.}, \textsc{Ridker,
  P.~M.} \& \textsc{Chasman, D.~I.} (2015).
\newblock Differential genetic effects on statin-induced changes across
  low-density lipoprotein--related measures.
\newblock \textit{Circulation: Cardiovascular Genetics} \textbf{8}, 688--695.

\bibitem[{DerSimonian(1996)}]{dersimonian1996meta}
\textsc{DerSimonian, R.} (1996).
\newblock Meta-analysis in the design and monitoring of clinical trials.
\newblock \textit{Statistics in medicine} \textbf{15}, 1237--1248.

\bibitem[{Doiron et~al.(2013)Doiron, Burton, Marcon, Gaye, Wolffenbuttel,
  Perola, Stolk, Foco, Minelli, Waldenberger et~al.}]{doiron2013data}
\textsc{Doiron, D.}, \textsc{Burton, P.}, \textsc{Marcon, Y.}, \textsc{Gaye,
  A.}, \textsc{Wolffenbuttel, B.~H.}, \textsc{Perola, M.}, \textsc{Stolk,
  R.~P.}, \textsc{Foco, L.}, \textsc{Minelli, C.}, \textsc{Waldenberger, M.}
  et~al. (2013).
\newblock Data harmonization and federated analysis of population-based
  studies: the {BioSHaRE} project.
\newblock \textit{Emerging themes in epidemiology} \textbf{10}, 12.

\bibitem[{Foster \& George(1994)}]{foster1994risk}
\textsc{Foster, D.~P.} \& \textsc{George, E.~I.} (1994).
\newblock The risk inflation criterion for multiple regression.
\newblock \textit{The Annals of Statistics} , 1947--1975.

\bibitem[{Franke et~al.(2010)Franke, McGovern, Barrett, Wang, Radford-Smith,
  Ahmad, Lees, Balschun, Lee, Roberts et~al.}]{franke2010genome}
\textsc{Franke, A.}, \textsc{McGovern, D.~P.}, \textsc{Barrett, J.~C.},
  \textsc{Wang, K.}, \textsc{Radford-Smith, G.~L.}, \textsc{Ahmad, T.},
  \textsc{Lees, C.~W.}, \textsc{Balschun, T.}, \textsc{Lee, J.},
  \textsc{Roberts, R.} et~al. (2010).
\newblock Genome-wide meta-analysis increases to 71 the number of confirmed
  crohn's disease susceptibility loci.
\newblock \textit{Nature genetics} \textbf{42}, 1118.

\bibitem[{Gaye et~al.(2014)Gaye, Marcon, Isaeva, LaFlamme, Turner, Jones,
  Minion, Boyd, Newby, Nuotio et~al.}]{gaye2014datashield}
\textsc{Gaye, A.}, \textsc{Marcon, Y.}, \textsc{Isaeva, J.}, \textsc{LaFlamme,
  P.}, \textsc{Turner, A.}, \textsc{Jones, E.~M.}, \textsc{Minion, J.},
  \textsc{Boyd, A.~W.}, \textsc{Newby, C.~J.}, \textsc{Nuotio, M.-L.} et~al.
  (2014).
\newblock {DataSHIELD}: taking the analysis to the data, not the data to the
  analysis.
\newblock \textit{International journal of epidemiology} \textbf{43},
  1929--1944.

\bibitem[{Huang \& Zhang(2010)}]{huang2010benefit}
\textsc{Huang, J.} \& \textsc{Zhang, T.} (2010).
\newblock The benefit of group sparsity.
\newblock \textit{The Annals of Statistics} \textbf{38}, 1978--2004.

\bibitem[{Jankov{\'a} \& Van De~Geer(2016)}]{jankova2016confidence}
\textsc{Jankov{\'a}, J.} \& \textsc{Van De~Geer, S.} (2016).
\newblock Confidence regions for high-dimensional generalized linear models
  under sparsity.
\newblock \textit{arXiv preprint arXiv:1610.01353} .

\bibitem[{Javanmard \& Montanari(2014)}]{javanmard2014confidence}
\textsc{Javanmard, A.} \& \textsc{Montanari, A.} (2014).
\newblock Confidence intervals and hypothesis testing for high-dimensional
  regression.
\newblock \textit{The Journal of Machine Learning Research} \textbf{15},
  2869--2909.

\bibitem[{Jones et~al.(2012)Jones, Sheehan, Masca, Wallace, Murtagh \&
  Burton}]{jones2012datashield}
\textsc{Jones, E.}, \textsc{Sheehan, N.}, \textsc{Masca, N.}, \textsc{Wallace,
  S.}, \textsc{Murtagh, M.} \& \textsc{Burton, P.} (2012).
\newblock {DataSHIELD}--shared individual-level analysis without sharing the
  data: a biostatistical perspective.
\newblock \textit{Norsk epidemiologi} \textbf{21}.

\bibitem[{Jordan et~al.(2019)Jordan, Lee \& Yang}]{jordan2018communication}
\textsc{Jordan, M.~I.}, \textsc{Lee, J.~D.} \& \textsc{Yang, Y.} (2019).
\newblock Communication-efficient distributed statistical inference.
\newblock \textit{Journal of the American Statistical Association}
  \textbf{526}, 668--681.

\bibitem[{Kozak \& Anunciado-Koza(2009)}]{kozak2009ucp1}
\textsc{Kozak, L.} \& \textsc{Anunciado-Koza, R.} (2009).
\newblock Ucp1: its involvement and utility in obesity.
\newblock \textit{International journal of obesity} \textbf{32}, S32.

\bibitem[{Lee et~al.(2017)Lee, Liu, Sun \& Taylor}]{lee2017communication}
\textsc{Lee, J.~D.}, \textsc{Liu, Q.}, \textsc{Sun, Y.} \& \textsc{Taylor,
  J.~E.} (2017).
\newblock Communication-efficient sparse regression.
\newblock \textit{Journal of Machine Learning Research} \textbf{18}, 1--30.

\bibitem[{Li et~al.(2016)Li, Liu, Yang \& Xie}]{li2016supporting}
\textsc{Li, W.}, \textsc{Liu, H.}, \textsc{Yang, P.} \& \textsc{Xie, W.}
  (2016).
\newblock Supporting regularized logistic regression privately and efficiently.
\newblock \textit{PloS one} \textbf{11}, e0156479.

\bibitem[{Liu \& Luo(2014)}]{liu2014hypo}
\textsc{Liu, W.} \& \textsc{Luo, S.} (2014).
\newblock Hypothesis testing for high-dimensional regression models.
\newblock \textit{Technical report} .

\bibitem[{Lounici et~al.(2011)Lounici, Pontil, Van De~Geer, Tsybakov
  et~al.}]{lounici2011oracle}
\textsc{Lounici, K.}, \textsc{Pontil, M.}, \textsc{Van De~Geer, S.},
  \textsc{Tsybakov, A.~B.} et~al. (2011).
\newblock Oracle inequalities and optimal inference under group sparsity.
\newblock \textit{The Annals of Statistics} \textbf{39}, 2164--2204.

\bibitem[{Ma et~al.(2020)Ma, Tony~Cai \& Li}]{ma2020global}
\textsc{Ma, R.}, \textsc{Tony~Cai, T.} \& \textsc{Li, H.} (2020).
\newblock Global and simultaneous hypothesis testing for high-dimensional
  logistic regression models.
\newblock \textit{Journal of the American Statistical Association} , 1--15.

\bibitem[{Mitra et~al.(2016)Mitra, Zhang et~al.}]{mitra2016benefit}
\textsc{Mitra, R.}, \textsc{Zhang, C.-H.} et~al. (2016).
\newblock The benefit of group sparsity in group inference with de-biased
  scaled group lasso.
\newblock \textit{Electronic Journal of Statistics} \textbf{10}, 1829--1873.

\bibitem[{Negahban et~al.(2012)Negahban, Ravikumar, Wainwright, Yu
  et~al.}]{negahban2012unified}
\textsc{Negahban, S.~N.}, \textsc{Ravikumar, P.}, \textsc{Wainwright, M.~J.},
  \textsc{Yu, B.} et~al. (2012).
\newblock A unified framework for high-dimensional analysis of $ m $-estimators
  with decomposable regularizers.
\newblock \textit{Statistical Science} \textbf{27}, 538--557.

\bibitem[{Nissen et~al.(2005)Nissen, Tuzcu, Schoenhagen, Crowe, Sasiela, Tsai,
  Orazem, Magorien, O'Shaughnessy \& Ganz}]{nissen2005statin}
\textsc{Nissen, S.~E.}, \textsc{Tuzcu, E.~M.}, \textsc{Schoenhagen, P.},
  \textsc{Crowe, T.}, \textsc{Sasiela, W.~J.}, \textsc{Tsai, J.},
  \textsc{Orazem, J.}, \textsc{Magorien, R.~D.}, \textsc{O'Shaughnessy, C.} \&
  \textsc{Ganz, P.} (2005).
\newblock Statin therapy, ldl cholesterol, c-reactive protein, and coronary
  artery disease.
\newblock \textit{New England Journal of Medicine} \textbf{352}, 29--38.

\bibitem[{Panagiotou et~al.(2013)Panagiotou, Willer, Hirschhorn \&
  Ioannidis}]{panagiotou2013power}
\textsc{Panagiotou, O.~A.}, \textsc{Willer, C.~J.}, \textsc{Hirschhorn, J.~N.}
  \& \textsc{Ioannidis, J.~P.} (2013).
\newblock The power of meta-analysis in genome-wide association studies.
\newblock \textit{Annual review of genomics and human genetics} \textbf{14},
  441--465.

\bibitem[{Rajpathak et~al.(2009)Rajpathak, Kumbhani, Crandall, Barzilai,
  Alderman \& Ridker}]{rajpathak2009statin}
\textsc{Rajpathak, S.~N.}, \textsc{Kumbhani, D.~J.}, \textsc{Crandall, J.},
  \textsc{Barzilai, N.}, \textsc{Alderman, M.} \& \textsc{Ridker, P.~M.}
  (2009).
\newblock Statin therapy and risk of developing type 2 diabetes: a
  meta-analysis.
\newblock \textit{Diabetes care} \textbf{32}, 1924--1929.

\bibitem[{Rodrigues et~al.(2013)Rodrigues, Sobrino, Genvigir, Willrich, Arazi,
  Dorea, Bernik, Bertolami, Faludi, Brion et~al.}]{rodrigues2013genetic}
\textsc{Rodrigues, A.~C.}, \textsc{Sobrino, B.}, \textsc{Genvigir, F. D.~V.},
  \textsc{Willrich, M. A.~V.}, \textsc{Arazi, S.~S.}, \textsc{Dorea, E.~L.},
  \textsc{Bernik, M. M.~S.}, \textsc{Bertolami, M.}, \textsc{Faludi, A.~A.},
  \textsc{Brion, M.} et~al. (2013).
\newblock Genetic variants in genes related to lipid metabolism and
  atherosclerosis, dyslipidemia and atorvastatin response.
\newblock \textit{Clinica Chimica Acta} \textbf{417}, 8--11.

\bibitem[{Stewart(2010)}]{stewart2010meta}
\textsc{Stewart, G.} (2010).
\newblock Meta-analysis in applied ecology.
\newblock \textit{Biology letters} \textbf{6}, 78--81.

\bibitem[{Study et~al.(2008)}]{study2008meta}
\textsc{Study, C.} et~al. (2008).
\newblock Meta-analysis of genome-wide association data identifies four new
  susceptibility loci for colorectal cancer.
\newblock \textit{Nature genetics} \textbf{40}, 1426.

\bibitem[{Swerdlow et~al.(2015)Swerdlow, Preiss, Kuchenbaecker, Holmes,
  Engmann, Shah, Sofat, Stender, Johnson, Scott et~al.}]{swerdlow2015hmg}
\textsc{Swerdlow, D.~I.}, \textsc{Preiss, D.}, \textsc{Kuchenbaecker, K.~B.},
  \textsc{Holmes, M.~V.}, \textsc{Engmann, J.~E.}, \textsc{Shah, T.},
  \textsc{Sofat, R.}, \textsc{Stender, S.}, \textsc{Johnson, P.~C.},
  \textsc{Scott, R.~A.} et~al. (2015).
\newblock Hmg-coenzyme a reductase inhibition, type 2 diabetes, and bodyweight:
  evidence from genetic analysis and randomised trials.
\newblock \textit{The Lancet} \textbf{385}, 351--361.

\bibitem[{Tang et~al.(2016)Tang, Zhou \& Song}]{tang2016method}
\textsc{Tang, L.}, \textsc{Zhou, L.} \& \textsc{Song, P. X.-K.} (2016).
\newblock Method of divide-and-combine in regularized generalized linear models
  for big data.
\newblock \textit{arXiv preprint arXiv:1611.06208} .

\bibitem[{Tong et~al.(2020)Tong, Duan, Li, Scheuemie, Moore \&
  Chen}]{tong2020robust}
\textsc{Tong, J.}, \textsc{Duan, R.}, \textsc{Li, R.}, \textsc{Scheuemie,
  M.~J.}, \textsc{Moore, J.~H.} \& \textsc{Chen, Y.} (2020).
\newblock Robust-odal: Learning from heterogeneous health systems without
  sharing patient-level data.
\newblock In \textit{Pacific Symposium on Biocomputing. Pacific Symposium on
  Biocomputing}, vol.~25. World Scientific.

\bibitem[{Van~de Geer et~al.(2014)Van~de Geer, B{\"u}hlmann, Ritov, Dezeure
  et~al.}]{van2014asymptotically}
\textsc{Van~de Geer, S.}, \textsc{B{\"u}hlmann, P.}, \textsc{Ritov, Y.},
  \textsc{Dezeure, R.} et~al. (2014).
\newblock On asymptotically optimal confidence regions and tests for
  high-dimensional models.
\newblock \textit{The Annals of Statistics} \textbf{42}, 1166--1202.

\bibitem[{Wang et~al.(2009)Wang, Li \& Leng}]{wang2009shrinkage}
\textsc{Wang, H.}, \textsc{Li, B.} \& \textsc{Leng, C.} (2009).
\newblock Shrinkage tuning parameter selection with a diverging number of
  parameters.
\newblock \textit{Journal of the Royal Statistical Society: Series B
  (Statistical Methodology)} \textbf{71}, 671--683.

\bibitem[{Wang et~al.(2017)Wang, Kolar, Srebro \& Zhang}]{wang2017efficient}
\textsc{Wang, J.}, \textsc{Kolar, M.}, \textsc{Srebro, N.} \& \textsc{Zhang,
  T.} (2017).
\newblock Efficient distributed learning with sparsity.
\newblock In \textit{Proceedings of the 34th International Conference on
  Machine Learning-Volume 70}. JMLR. org.

\bibitem[{Waters et~al.(2013)Waters, Ho, Boekholdt, DeMicco, Kastelein, Messig,
  Breazna \& Pedersen}]{waters2013cardiovascular}
\textsc{Waters, D.~D.}, \textsc{Ho, J.~E.}, \textsc{Boekholdt, S.~M.},
  \textsc{DeMicco, D.~A.}, \textsc{Kastelein, J.~J.}, \textsc{Messig, M.},
  \textsc{Breazna, A.} \& \textsc{Pedersen, T.~R.} (2013).
\newblock Cardiovascular event reduction versus new-onset diabetes during
  atorvastatin therapy: effect of baseline risk factors for diabetes.
\newblock \textit{Journal of the American College of Cardiology} \textbf{61},
  148--152.

\bibitem[{Wolfson et~al.(2010)Wolfson, Wallace, Masca, Rowe, Sheehan, Ferretti,
  LaFlamme, Tobin, Macleod, Little et~al.}]{wolfson2010datashield}
\textsc{Wolfson, M.}, \textsc{Wallace, S.~E.}, \textsc{Masca, N.},
  \textsc{Rowe, G.}, \textsc{Sheehan, N.~A.}, \textsc{Ferretti, V.},
  \textsc{LaFlamme, P.}, \textsc{Tobin, M.~D.}, \textsc{Macleod, J.},
  \textsc{Little, J.} et~al. (2010).
\newblock {DataSHIELD}: resolving a conflict in contemporary
  bioscience???performing a pooled analysis of individual-level data without
  sharing the data.
\newblock \textit{International journal of epidemiology} \textbf{39},
  1372--1382.

\bibitem[{Xia et~al.(2018{\natexlab{a}})Xia, Cai \& Cai}]{xia2018two}
\textsc{Xia, Y.}, \textsc{Cai, T.} \& \textsc{Cai, T.~T.} (2018{\natexlab{a}}).
\newblock Two-sample tests for high-dimensional linear regression with an
  application to detecting interactions.
\newblock \textit{Statistica Sinica} \textbf{28}, 63.

\bibitem[{Xia et~al.(2018{\natexlab{b}})Xia, Cai \& Li}]{xia2018joint}
\textsc{Xia, Y.}, \textsc{Cai, T.~T.} \& \textsc{Li, H.} (2018{\natexlab{b}}).
\newblock Joint testing and false discovery rate control in high-dimensional
  multivariate regression.
\newblock \textit{Biometrika} \textbf{105}, 249--269.

\bibitem[{Zeggini et~al.(2008)Zeggini, Scott, Saxena, Voight, Marchini, Hu,
  de~Bakker, Abecasis, Almgren, Andersen et~al.}]{zeggini2008meta}
\textsc{Zeggini, E.}, \textsc{Scott, L.~J.}, \textsc{Saxena, R.},
  \textsc{Voight, B.~F.}, \textsc{Marchini, J.~L.}, \textsc{Hu, T.},
  \textsc{de~Bakker, P.~I.}, \textsc{Abecasis, G.~R.}, \textsc{Almgren, P.},
  \textsc{Andersen, G.} et~al. (2008).
\newblock Meta-analysis of genome-wide association data and large-scale
  replication identifies additional susceptibility loci for type 2 diabetes.
\newblock \textit{Nature genetics} \textbf{40}, 638.

\bibitem[{Zhang \& Zhang(2014)}]{zhang2014confidence}
\textsc{Zhang, C.-H.} \& \textsc{Zhang, S.~S.} (2014).
\newblock Confidence intervals for low dimensional parameters in high
  dimensional linear models.
\newblock \textit{Journal of the Royal Statistical Society: Series B
  (Statistical Methodology)} \textbf{76}, 217--242.

\bibitem[{Z{\"o}ller et~al.(2018)Z{\"o}ller, Lenz \&
  Binder}]{zoller2018distributed}
\textsc{Z{\"o}ller, D.}, \textsc{Lenz, S.} \& \textsc{Binder, H.} (2018).
\newblock Distributed multivariable modeling for signature development under
  data protection constraints.
\newblock \textit{arXiv preprint arXiv:1803.00422} .

\end{thebibliography}



\end{document}